\documentclass[12pt]{article}
\usepackage[utf8]{inputenc}
\usepackage[T1]{fontenc}
\usepackage{amsmath}
\usepackage{graphicx}
\usepackage{psfrag,epsf}
\usepackage{enumerate}
\usepackage{natbib}
\usepackage{url} 

\usepackage{tikz}
\usepackage{enumitem}
\usepackage{multicol}
\usepackage{amsfonts}
\usepackage{amssymb}

\newcommand{\blind}{0}

\newcommand{\indep}{\rotatebox[origin=c]{90}{$\models$}}

\begin{document}

\def\spacingset#1{\renewcommand{\baselinestretch}%
{#1}\small\normalsize} \spacingset{1}


\if0\blind
{
  \title{\bf Hypothetical estimands in clinical trials: \\ a unification of causal inference and missing data methods}
  \author{Camila Olarte Parra\thanks{
    This work was funded by a UK Medical Research Council grant (MR/T023953/1)}\hspace{.2cm}\\
    Department of Mathematical Sciences, University of Bath\\
    Rhian M. Daniel \\
    Division of Population Medicine, Cardiff University \\
    and \\
    Jonathan W. Bartlett \\
    Department of Mathematical Sciences, University of Bath}
  \maketitle
} \fi

\if1\blind
{
  \bigskip
  \bigskip
  \bigskip
  \begin{center}
    {\LARGE\bf Title}
\end{center}
  \medskip
} \fi

\bigskip
\begin{abstract}
The ICH E9 addendum introduces the term intercurrent event to refer to events that happen after randomisation and that can either preclude observation of the outcome of interest or affect its interpretation. It proposes five strategies for handling intercurrent events to form an estimand but does not suggest statistical methods for estimation. In this paper we focus on the hypothetical strategy, where the treatment effect is defined under the hypothetical scenario in which the intercurrent event is prevented. For its estimation, we consider causal inference and missing data methods. We establish that certain ‘causal inference estimators’ are identical to certain ‘missing data estimators’. These links may help those familiar with one set of methods but not the other. Moreover, using potential outcome notation allows us to state more clearly the assumptions on which missing data methods rely to estimate hypothetical estimands. This helps to indicate whether estimating a hypothetical estimand is reasonable, and what data should be used in the analysis. We show that hypothetical estimands can be estimated by exploiting data after intercurrent event occurrence, which is typically not used. We also present Monte Carlo simulations that illustrate the implementation and performance of the methods in different settings. 
\end{abstract}

\noindent%
\vfill

\newpage
\spacingset{1.45} 
\section{Introduction}
\label{sec:intro}
The analysis of randomised trials is often complicated by the occurrence of certain events that affect the interpretation of the treatment effect or preclude the observation of the outcome of interest. Such occurrences, termed `intercurrent events' (ICE) by the recently-published ICH E9 addendum on estimands \citep{ICHE9Addendum}, include treatment discontinuation, addition of rescue medication, or death prior to measurement of the outcome of interest. In the presence of such ICEs, the importance of clear specification of a trial's treatment effect `estimand' and how the statistical analysis targets this estimand has been increasingly recognised over the last decade. 

The US National Research Council report on the Prevention and Handling of Missing Data in Clinical Trials highlighted the importance of trials clearly specifying the target estimand(s), and how the trial design and statistical analysis should be chosen to support its reliable estimation \citep{national2010prevention}. Since then, a number of authors have considered the complex questions involved in how to choose and specify an estimand and how to select an appropriate statistical method to estimate it \citep{mallinckrodt2012structured,Mallinckrodt2019EstimandBook,mallinckrodt2020aligning,carpenter2013analysis,holzhauer2015choice}.

According to the ICH E9 framework, strategies for dealing with intercurrent events must be specified when choosing and defining the target estimand of a clinical trial. The addendum does not however specify how these might or should correspond to statistical analysis methods. One of the proposed strategies is labelled as hypothetical. Under the hypothetical strategy, the causal effect is targeting what would have happened if the ICE had (somehow) been prevented from occurring. For patients in the trial for whom the ICE did not occur, their observed outcome corresponds to the outcome of interest under the hypothetical strategy, whereas for those who experienced the ICE, the potential outcome of interest is missing. Consequently, the existing literature has almost exclusively focused on tackling the problem of estimation of hypothetical estimands from the perspective of missing data, by deleting any outcomes observed after ICE occurrence and applying methods such as direct likelihood (e.g.\ using linear mixed models) or multiple imputation.

Until recently \citep{lipkovich2020causal,Bowden2020IV,Michiels2021Balanced,Qu2021PrincipalStrat}, relatively little has been published on the topic of estimation of estimands from the perspective of modern casual inference. Indeed, perhaps surprisingly, the ICH E9 addendum itself does not explicitly mention causal inference concepts or methods, although these are clearly relevant.

The hypothetical strategy has been used when the ICE is addition of rescue medication. A recent systematic review on rescue medication in trials of asthma and eczema found that its use was not routinely reported or accounted for, even when there was evidence of an imbalance in rescue medication between arms \citep{Ster2020SystermaticReview}. When analyses aiming to account for rescue medication were reported, the authors of the review considered that they were mainly targeting a hypothetical estimand with suboptimal methods and concluded that further guidance was warranted.

In diabetes trials, rescue medication is usually available for ethical reasons. An example of such a trial compared dapagliflozin, dapagliflozin plus saxagliptin, and glimepiride in patients with type 2 diabetes who were using metformin \citep{Muller2018Diabetes}. Insulin therapy was available as rescue medication. The analysis of the primary end point was performed using a linear mixed model with fixed effects for treatment group and covariates after deleting values beyond the first use of rescue medication. Here, we will discuss alternative approaches and their corresponding underlying assumptions, using either only the values prior to rescue medication or the full observed values even after the ICE and how they relate to this approach. 



In this paper, we review concepts from causal inference to characterise precisely the conditions under which hypothetical estimands can be estimated from trial data. We describe statistical estimators of these arising from both the causal inference and missing data literatures, and establish that for each missing data estimator there is a corresponding numerically identical causal inference estimator, thereby unifying the sets of methods.

We begin in Section \ref{sec:causalintro} with a review of the concepts and tools in causal inference, first for a setting with a treatment decision at a single time point, and then for a generalised setting where treatment changes can occur at multiple times. In Section \ref{sec:singletime} we consider the definition and estimation of a hypothetical estimand in a simplified setting in which the ICE can only occur at a single time point, linking it to the concepts and methods reviewed in Section \ref{sec:causalintro}. In Section \ref{sec:multipletime} we consider the more general setup in which an ICE can occur at multiple time points. Using Monte Carlo simulations, we explore in Section \ref{sec:sims} the validity and efficiency of these methods to account for ICE in different settings. Finally, we give conclusions in Section \ref{sec:conc}.

\section{A brief review of causal inference concepts, assumptions, and estimators}
\label{sec:causalintro}
In this section we review the key concepts, assumptions and estimation methods from causal inference for studies with time-varying treatments, drawing on \cite{Hernan2020}, \cite{Robins2009Chapter}, \cite{tsiatis2020dynamic}, and \cite{Daniel2013}. 

\subsection{Time-fixed treatment}
Clinical trials usually compare two or more treatments for a given condition and evaluate their effects on an outcome of interest. The potential outcomes framework provides a formal definition for such causal effects and the assumptions required to estimate them \citep{Rubin1974}. `Potential outcome' refers to the response that would have been observed on a patient had they been assigned a particular treatment. Thus, there is a potential outcome for each patient for every treatment we might feasibly assign to them. Except in certain special situations, patients only receive one treatment and therefore only one of their potential outcomes is observed. 

The potential outcome, denoted $Y^a$, expresses the outcome $Y$ under assignment to treatment $a$. We can then define the target causal effect of interest (the estimand) as a contrast of the distributions of such potential outcomes. For a dichotomous treatment, $A$, we may for example be interested in the mean difference, $E(Y^{a=1}) - E(Y^{a=0})$, or simply $E(Y^1) - E(Y^0)$.

In RCTs, the treatment at baseline is assigned at random but the occurrence of the ICE is not. As we will see in the next section, the ICE can be considered a treatment that is not randomly assigned. The causal effect of interest can be estimated if certain 
\textit{identifiability assumptions} are satisfied. First, the interventions have to be sufficiently well defined to ensure \textit{consistency}, which states that the observed outcome corresponds to the potential outcome under the treatment received:  $Y=Y^a$ if $A=a$, where $A$ denotes the variable recording the treatment a given patient receives \citep{Vanderweele2009Consistency}. Consider an oncology trial where we compare chemotherapy vs no chemotherapy. The chemotherapy treatment would be considered ill-defined if the type of chemotherapy and regimen are not specified. Also no chemotherapy can imply no treatment or follow up at all or just standard of care or many other options. As we would not expect that different types of chemotherapy regimens would yield similar outcomes, then the potential outcomes $Y^{a}$ are not sufficiently well defined.

As patients can usually receive only one treatment, we compare different groups of patients receiving the different treatments of interest. Randomisation ensures that the different groups of patients have similar prognostic factor distributions. In the absence of randomisation, the effect estimate has to be adjusted for a sufficient set of confounders to ensure that patients are comparable in terms of their prognostic factors. This is the second identifiability condition known as \textit{conditional exchangeability}. In other words, each of the potential outcomes $Y^{a}$ for the different possible values of $a$ has to be independent of the actual treatment received $A$, given the confounders $L$: $Y^a \indep A|L$. 

Finally, the last identifiability assumption is \textit{positivity}. This means that for every patient, on the basis of their confounder values $L$, there is a non-zero probability that they could receive each of the treatments under study \citep{Petersen2012Positivity}. It would not be sensible to consider patients who, on the basis of one or more of their confounder values, would always receive a given treatment. This could happen if, say, a particular confounder level implies contraindication for one of the treatments. Therefore, all patients should have a non-zero probability of receiving the different treatments $ P(A=a|L=l) > 0$ for all values of $a$ and $l$ such that $P(L=l)>0$.

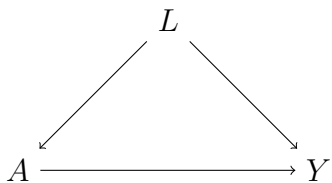
\begin{figure}[hbt!]
    \centering
    \begin{tikzpicture}
    \node (A) at (0.00,0.00) {$A$};
    \node (L) at (2.00,2.00) {$L$};
    \node (Y) at (4.00,0.00) {$Y$};

    \draw [->] (L) edge (A);
    \draw [->] (L) edge (Y);
    \draw [->] (A) edge (Y);
    \end{tikzpicture}
    \caption{Example of a direct acyclic graphs (DAG) relating treatment $A$, outcome $Y$ and confounders $L$}
    \label{fig:DAG_example}
\end{figure}

A useful way to encode causal assumptions is to use so-called directed acyclic graphs (DAG). These graphs are composed of nodes that represent random variables, including treatment, outcome and confounders, and the edges connecting the nodes (Figure \ref{fig:DAG_example}). DAGs are said to be directed because the edges have arrows indicating the direction of the causal effect and acyclic because a directed path i.e.\ edge or series of consecutive edges, cannot lead back to an initial node. The absence of an edge between two nodes encodes the assumption that there is no direct causal effect between them. In Figure \ref{fig:DAG_example}, we could omit the edge $L \rightarrow A$ in an RCT to show that $A$ is assigned at random.

\subsection{Time-varying treatment}
\label{review:timevarying}
\begin{figure}[hbt!]
    \centering
    \begin{tikzpicture}
    \node (A_0) at (0.00,0.00) {$A_0$};
    \node (A_1) at (3.00,0.00) {$A_1$};
    \node (A_2) at (6.00,0.00) {$A_2$};
    \node (L_0) at (-1.00,2.00) {$L_0$};
    \node (L_1) at (2.00,2.00) {$L_1$};
    \node (L_2) at (5.00,2.00) {$L_2$};
    \node (Y) at (9.00,0.00) {$Y$};
    
    \draw [->] (L_0) edge (L_1);
    \draw [->] (L_0) edge (A_0);
    \draw [->] (L_0) edge (A_1);
    \draw [->] (L_0) edge (A_2);
    \draw [->] (L_0) to[out=90,in=90] (L_2);
    \draw [->] (L_0) to[out=90,in=90] (Y);
    
    \draw [->] (A_0) edge (A_1);
    \draw [->] (A_0) to[out=-90,in=-90] (A_2);
    \draw [->] (A_0) to[out=-90,in=-90] (Y);
    \draw [->] (A_0) edge (L_1);
    \draw [->] (A_0) to[out=90,in=90] (L_2);
    
    \draw [->] (A_1) edge (A_2);
    \draw [->] (A_1) to[out=-90,in=-90] (Y);
    \draw [->] (A_1) edge (L_2);
    
    \draw [->] (A_2) edge (Y);
    
    \draw [->] (L_1) edge (L_2);
    \draw [->] (L_1) edge (A_1);
    \draw [->] (L_1) edge (A_2);
    \draw [->] (L_1) edge (Y);
    
    \draw [->] (L_2) edge (A_2);
    \draw [->] (L_2) edge (Y);
    
    \end{tikzpicture}

    \caption{Directed acyclic graph (DAG) of a study with time-varying treatment}
    \label{fig:causalintroDAG_threetimes}
\end{figure}
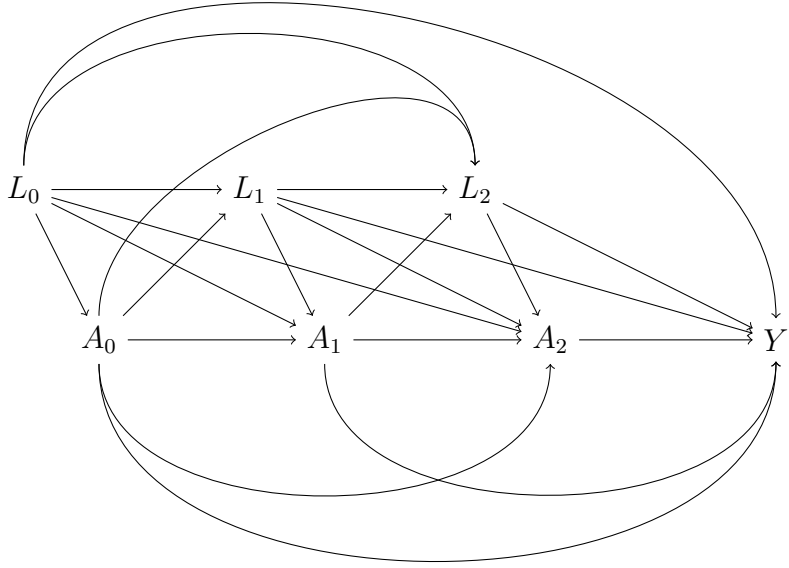

The preceding setup can be extended to a more general one where treatment can change over time. Figure \ref{fig:causalintroDAG_threetimes} shows a possible DAG for a study where treatment can change at three time points, with $A_0,A_1,A_2$ denoting variables for treatment at each time. The treatment at a given time point $k$ can depend on the earlier treatment values and also the earlier values of the variables $L$. The final outcome of interest is denoted $Y$.

In the setting with time-varying treatments we are typically interested in comparing different treatment regimes, that is different (hypothetical) ways of assigning treatments. A static treatment regime is one in which the decision on which treatment to assign at each time point does \textit{not} depend on the time-varying variables $L$, whereas dynamic regimes are those where the treatment decisions can be based on the hypothetical values of $L$ that would be observed under that regime. As we will see later, the hypothetical estimand of `no ICE' corresponds to a static regime. For the DAG in Figure \ref{fig:causalintroDAG_threetimes} a static regime is defined by specifying particular values for the three treatment variables. For example, assuming there are two treatments available, coded 0 and 1, a particular regime is $\bar{a}=(0,0,0)$, which corresponds to assigning the first treatment at each of the three time points. The potential outcome under this regime is denoted $Y^{\bar{a}}$.

The identifiability conditions described previously for the setting with a single treatment assignment can be extended to this more general setting with time-varying treatments. The no unmeasured confounding or exchangeability condition has a number of different versions in the time-varying treatment setting. For our purposes we use the following version, which states
\begin{equation}
    Y{^{\bar{a}}} \indep A_k | \bar{A}_{k-1}=\bar{a}_{k-1}, \bar{L}_k
    \label{eq:seq_exch_ass}
\end{equation}
for $k=0,1,\dots,K$ and all static treatment regimes $\bar{a}$, where  $\bar{A}_{k-1}=(A_{0},\dots,A_{k-1})$ denotes the history of treatment received through to time $k-1$, and similarly for $\bar{a}_{k-1}$ and $\bar{L}_{k}$. This condition is satisfied if treatment assignment at each time point depends only on previous treatments and measured time-varying confounders, as in the DAG in Figure \ref{fig:causalintroDAG_threetimes}. We will see later that this assumption plays a critical role regarding which variables are included in $L_k$ in order to provide valid inferences.

The positivity condition is similarly extended in the time-varying treatment setting to the following
\begin{equation}
  P(A_k=a_k | \bar{A}_{k-1}=\bar{a}_{k-1}, \bar{L}_k=\bar{l}_k) > 0 \text{ for all } (a_k, \bar{a}_{k-1}, \bar{l}_k)   \text{ that satisfy } f_{\bar{A}_{k-1},\bar{L}_{k}}(\bar{a}_{k-1},\bar{l}_{k}) > 0 
    \label{eq:positivity_ass}
\end{equation}
for each $k$. In words, this says that for all combinations of treatment histories and time-varying confounders up to time $k-1$ that can occur in the study, there is a positive probability of each of the possible treatments being given at time $k$. In fact, if we are interested in a particular treatment strategy, this condition only needs to hold for treatment histories compatible with the strategy of interest (see Technical Point 19.2 of \citet{Hernan2020}). For example, if we are interested in the treatment strategy of giving treatment zero throughout, $\bar{a}=(0,\dots,0)$, we require only that
\begin{equation*}
 P(A_k=0 | \bar{A}_{k-1}=\bar{0}_{k-1}, \bar{L}_k=\bar{l}_k) > 0 \text{ for all } \bar{l}_k \text{ that satisfy }  f_{\bar{A}_{k-1},\bar{L}_{k}}(\bar{0}_{k-1},\bar{l}_{k}) > 0 
\end{equation*}
for each $k$. We will see later that the positivity assumption plays an important role in the feasibility of estimation of hypothetical estimands where whether an ICE occurs depends deterministically on time-varying confounders. 

In what follows, where there is no ambiguity introduced, we use $f(x|a,b,c)$ as shorthand for $f_{X|A,B,C}(x|a,b,c)$. In the case that $X$ is discrete, $f(x|a,b,c) = P(X=x|A=a,B=b,C=c)$.

\subsection{G-formula}\label{G-formula_intro}

We now review the two most commonly adopted approaches for estimation of the causal effects of treatment in the time-varying treatment setting. The first is G-formula or G-computation. Under the previously stated identification conditions, for a given treatment regime $\bar{a}$ the density function of the potential outcomes under this regime can be shown (Section 5.4 of \cite{tsiatis2020dynamic}) to be equal to
\begin{align}
    f_{Y^{\bar{a}}}(y) = \int_{l_0} \int_{l_1} \int_{l_2} f_{Y|\bar{A},\bar{L}}(y|\bar{a},\bar{l}) f_{L_2|\bar{A}_1,\bar{L}_1}(l_2|\bar{a}_{1},\bar{l}_{1}) f_{L_1|A_0,L_0}(l_1|a_0,l_0) f_{L_0}(l_0) dl_2 dl_1 dl_{0}
    \label{eq:gform_timevar_dens}
\end{align}
where we have taken $K=2$ for concreteness, $\bar{a}=(a_0,a_1,a_2)$ and $\bar{a}_1=(a_0,a_1)$. Often we will be interested in the mean outcome under a given treatment regime, which can then be shown to equal
\begin{align}
     E(Y^{\bar{a}}) &=  \int_{l_0} \int_{l_1} \int_{l_2} E(Y|\bar{a},\bar{l})  f(l_2|\bar{a}_1,\bar{l}_1)f(l_1|a_0,l_0) f(l_0) dl_2 dl_1 dl_0 \nonumber \\
    &=E\left( E\left[ E\left\{ E(Y|\bar{A}=\bar{a},\bar{L}) | \bar{A}_1=\bar{a}_1, \bar{L}_{1} \right\} | A_0=a_0,L_0 \right]  \right) \label{eq:gform_timevar_exp}
\end{align}
To implement G-formula we can specify and fit models for the conditional distributions
\begin{align}
    & f(y|\bar{a},\bar{l}), \nonumber \\
    & f(l_{k}|\bar{a}_{k-1},\bar{l}_{k-1}), \quad k=1,\dots,K, \nonumber \\
    & f(l_0)
    \label{eq:gformModelsNeeded}
\end{align}
If we are only interested in the mean outcome (as opposed to other aspects of the distribution) under the treatment strategy, then the conditional model for $Y$ can be replaced with a model for its conditional expectation. Also, taking a final simple average after all but the outer integral has been evaluated circumvents the need to specify a distribution for $L_0$. In general, the full conditional distributions of the covariates given their history must be specified even when only the mean of $Y^{\bar{a}}$ is of interest. Depending on the form of the model specified for $E(Y|\bar{a},\bar{l})$, however, only the implied lower moments of these distributions may be involved. As described in Section 5.5.1 of \cite{tsiatis2020dynamic}, when the time-varying confounders $L_k$ are univariate, this is relatively straightforward (although potentially expensive in terms of parametric assumptions), as each model is a univariate regression model. When the $L_k$ are multivariate, these regressions become multivariate, which is more difficult, particularly if the components of $L_k$ are a mixture of continuous and discrete variables. An alternative is to factorise $f(l_{k}|\bar{a}_{k-1},\bar{l}_{k-1})$ into a product of univariate conditional distributions, with each univariate regression model chosen according to the type of (scalar) variable. An issue for this approach is that it is not usually clear which order should be chosen. We note that a similar issue arises in the specification of imputation models for multivariate missing data \citep{erler2019jointai}.

Having specified and fitted the models, the G-formula identification equations \eqref{eq:gform_timevar_dens} or \eqref{eq:gform_timevar_exp} can be used. However, evaluation of the integrals involved in general is difficult. To circumvent this, a Monte-Carlo integration approach can be used, in which the values of the time-varying confounders $L_k$ are simulated from the fitted models sequentially. For further details, see \cite{Daniel2011GformulaStata}.

The resulting G-formula estimates of $E(Y^{\bar{a}})$ are consistent provided the previously stated identification assumptions hold and the models for the conditional distributions of the time-varying confounders and the model for the outcome \eqref{eq:gformModelsNeeded}, are correctly specified.

\subsection{Inverse probability of treatment weighting}
\label{review_ipw}
A different approach is to use inverse probability of treatment weighting to create a pseudopopulation in which the time-varying treatment assignment is independent of the values of the (time-varying) covariates. In other words, we create a pseudopopulation in which there are no longer arrows into the treatment nodes from any other node, but all other relationships remain unaltered. This is achieved by weighting each patient by the inverse of the probability of receiving the treatment they in fact received at each time point given the covariate and treatment history. This is the time-varying treatment extension of the propensity score (the probability of treatment given covariates). In the time-varying treatment setting, the (unstabilised) weight for patient $i$ is:
\begin{equation}
    W_{i} = \prod_{k=0}^{K} \frac{1}{f(A_{k,i}|\bar{A}_{k-1,i},\bar{L}_{k,i})}
    \label{eq:ipwGeneralUnstabilised}
\end{equation}
where patient $i$'s treatment history is $(A_{0,i},A_{1,i},\dots,A_{K,i})$, their time-varying confounder history is $(L_{0,i},L_{1,i},\dots,L_{K,i})$, and $\bar{A}_{-1,i}$ is taken to be zero for all $i$ by definition. The potential mean outcome under the treatment strategy of interest can then be estimated as the weighted average of the outcomes among those patients whose treatment history matches the treatment strategy $\bar{a}$ we are targeting:
\begin{equation}
    \frac{\sum^{n}_{i=1} I(\bar{A}_{i}=\bar{a}) W_{i} Y_{i}}{\sum^{n}_{i=1} I(\bar{A}_{i}=\bar{a}) W_{i}}
    \label{eq:ipw_gen_timevarying}
\end{equation}
where $I(\cdot)$ denotes the indicator function.

In practice the conditional distributions of the time-varying treatment variables involved in the definition of the weights are not known, but must instead be estimated. When there are only two treatment options at each time, such that each $A_k$ is binary, these can be logistic regression models. The resulting estimator of $E(Y^{\bar{a}})$ is consistent provided these models are correctly specified, along with the same identification assumptions stated previously. 

\section{Definition and estimation of a hypothetical estimand in a simplified setting}
\label{sec:singletime}
In this section we apply the assumptions and methods introduced in Section \ref{sec:causalintro} to the problem of estimating a hypothetical estimand in a randomised trial in which the ICE can take place at only a single fixed time point. We then compare and contrast the causal inference estimators with missing data estimators which are currently more commonly adopted in practice for the estimation of hypothetical estimands.

\begin{figure}[hbt!]
\centering
    \begin{tikzpicture}
    \node (A_0) at (0.00,0.00) {$A_0$};
    \node (A_1) at (3.00,0.00) {$A_1$};
    \node (L_0) at (-1.00,2.00) {$L_0$};
    \node (L_1) at (2.00,2.00) {$L_1$};
    \node (Y) at (6.00,0.00) {$Y$};
    
    \draw [->] (L_0) edge (L_1);
    \draw [->] (L_0) edge (A_1);
    \draw [->] (L_0) to[out=90,in=90] (Y);
    
    \draw [->] (A_0) edge (A_1);
    \draw [->] (A_0) to[out=-90,in=-90] (Y);
    \draw [->] (A_0) edge (L_1);
    
    \draw [->] (A_1) edge (Y);
    
    \draw [->] (L_1) edge (A_1);
    \draw [->] (L_1) edge (Y);
    
    \end{tikzpicture}
    
    \caption{Directed acyclic graph (DAG) representation of a randomised trial where the ICE occurring can only occur at a fixed time point. $A_0$ denotes randomised treatment, $A_1$ occurrence of ICE, $L_0$ baseline variables, $L_1$ post-baseline variables measured before the occurence of the ICE and $Y$ final outcome.}
    \label{fig:DAG_fixed}
\end{figure}
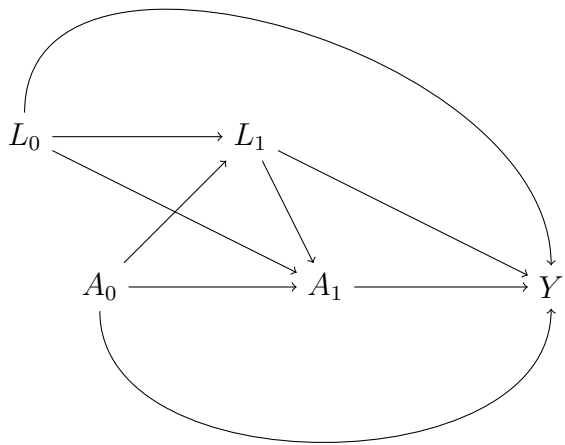

We consider inference under the setup depicted by the DAG in Figure \ref{fig:DAG_fixed}, which is a special case of the general time-varying treatment setting described in Section \ref{review:timevarying}. Here $A_0$ represents randomised treatment at baseline. Unlike Figure \ref{fig:causalintroDAG_threetimes}, there is no arrow from $L_0$ to $A_0$ due to the fact that treatment at baseline is randomly assigned. The second `treatment' variable $A_1$ represents whether or not the ICE occurs for a particular patient.

The potential outcomes $Y^{a_0,a_1}$ denote the outcome were we to assign treatment $a_0$ and intervene (somehow) on the ICE to set it to level $a_1$. The effect of treatment 1 vs.\ 0 in the hypothetical scenario were the ICE  prevented from occurring is thus:
\begin{equation}
    E(Y^{1,0}) - E(Y^{0,0})
    \label{eq:hyp_def}
\end{equation}

In some cases, the post-baseline variables ($L_1$) can include measurements of the outcome at intermediate visits. For instance in a diabetes trial, $L_1$ might consist of measurements of glycemic control, like glycated haemoglobin (HbA1c) and fasting plasma glucose (FPG), assessed at an intermediate follow-up visit, while $Y$ denotes HbA1c at the second (final) visit.

\subsection{Causal inference estimation approaches}
The sequential exchangeability assumption (equation \eqref{eq:seq_exch_ass}) holds here, given covariates $L_0,L_1$, since the DAG in Figure \ref{fig:DAG_fixed} is a special case of the general DAG in Figure \ref{fig:causalintroDAG_threetimes}. Informally this can be seen to hold here because the treatment $A_0$ is randomly assigned and the ICE $A_1$ (which is analogous to the `second treatment') is only influenced by assigned treatment $A_0$ and the measured variables $L_0$ and $L_1$. Thus $Y^{a_0,a_1=0} \indep A_0$ and $Y^{a_0,a_1=0} \indep A_1 | A_0, L_0, L_1$. We emphasize that $L_0$ and $L_1$ consist of all variables which influence the ICE occurrence ($A_1$) and the outcome $Y$. In particular, it is not sufficient to specify $L_0$ and $L_1$ as simply the baseline and intermediate measures of the outcome variable unless these are truly the only variables influencing the occurrence of the ICE and outcome.

For the positivity assumption (equation \eqref{eq:positivity_ass}), first since in randomised trials $P(A_0=0)=P(A_0=1)=0.5$, it is clearly the case that $P(A_0=a_0|L_0=l_0)=P(A_0=a_0)>0$ for both $a_0=0$ and $a_0=1$. Second, the occurrence of the ICE should also satisfy that $P(A_1=0|A_0=a_{0},L_0=l_{0},L_1=l_{1})>0$ for all possible $l_{0}$ and $l_{1}$ values in both treatment arms. This means that for all possible combinations of $L_0$, $L_1$ and $A_0$, the probability of \emph{not} having the ICE must be non-zero. Note that as we are only interested in the potential outcomes in the abscence of the ICE, then $P(A_1=1|A_0=a_{0},L_0=l_{0},L_1=l_{1})>0$ is not required, which implies that for some values of $a_{0}$, $l_{0}$ and $l_{1}$ it can be that $P(A_1=0|A_0=a_{0},L_0=l_{0},L_1=l_{1})=1$ without representing a violation of the positivity assumption.

\subsubsection{G-formula}
From equation \eqref{eq:gform_timevar_exp} the G-formula estimator for the mean potential outcome under a general treatment regime $\bar{a}$ is given by
\begin{equation*}
    E(Y^{\bar{a}}) = \int_{l_0} \int_{l_1} E(Y|\bar{a}, \bar{l}) f (l_1|a_0,l_0) f(l_0)  \,dl_1 \,dl_0
\end{equation*}
Unlike in the general observational study setting, in our setting the first treatment $A_0$ is randomly assigned. This means in particular that $A_0 \indep L_0$, which in turn means that $f(l_0) = f(l_0|a_0)$. Thus the preceding equation can in our setting be written as
\begin{align*}
    E(Y^{\bar{a}}) &= \int_{l_0} \int_{l_1} E(Y|\bar{a}, \bar{l}) f (l_1|a_0,l_0) f(l_0|a_0)  \,dl_1 \,dl_0 \\
    &= \int_{l_0} \int_{l_1} E(Y|\bar{a}, \bar{l}) f(l_1,l_0|a_0) \,dl_1 \,dl_0 
\end{align*}
To use this to estimate $E(Y^{\bar{a}})$ we could specify models for $E(Y|\bar{a}, \bar{l})$ and $f(l_1,l_0|a_0)$. However, in the interests of robustness we can instead `model' the latter joint distribution non-parametrically based on the empirical distribution of $L_0$ and $L_{1}$ among those randomised to $A_{0}=a_{0}$. This motivates the G-formula estimator:
\begin{align}
     \widehat{E}(Y^{\bar{a}}) &= \frac{ \sum^{n}_{i=1} I(A_{0,i}=a_0) \widehat{E}(Y_{i}|a_0, a_1, L_{0,i}, L_{1,i}) }{ \sum^{n}_{i=1} I(A_{0,i}=a_0)}
     \label{gformula1}
\end{align}
which relies only on a model for $E(Y|\bar{a},\bar{l})$, i.e.\ an appropriate model for the mean of $Y$ with $A_0$, $A_1$, $L_0$ and $L_1$ as covariates, noting that our estimator then requires predictions from this model where $A_1$ is set to $a_1$. If $Y$ is continuous we might for example choose a linear regression model with main effects of randomised treatment $A_0$, occurrence of ICE $A_1$, $L_0$ and $L_1$. This model would make use of all the observed data, including outcomes $Y$ which take place after an ICE. Through the inclusion of $A_1$ as a covariate, it models how the ICE influences the final outcome $Y$. When the ICE is receipt of rescue medication, this corresponds to use of the post-rescue outcomes with adjustment for rescue, and this type of approach has been discussed previously by \cite{holzhauer2015choice}.

As described previously, the hypothetical estimand corresponds to the contrast of the treatment regimes $\bar{a}=(1,0)$ and $\bar{a}=(0,0)$, where we set $A_1$ to $0$. A G-formula estimator for the hypothetical estimand under no ICE is thus
\begin{align}
     \widehat{E}(Y^{a_{0},a_1=0}) &= \frac{ \sum^{n}_{i=1} I(A_{0,i}=a_0) \widehat{E}(Y_{i}|a_0, a_1=0, L_{0,i}, L_{1,i}) }{ \sum^{n}_{i=1} I(A_{0,i}=a_0)}
     \label{gformulaNoICEGeneral}
\end{align}
The G-formula estimator is thus predicting, for every patient randomised to treatment group $a_0$, what their outcome would be were the ICE set to not occur.

Since here we are only interested in regimes which set $a_1=0$, we do not need in fact to model how occurrence of the ICE influences $Y$, since we only need to predict outcomes under no ICE $a_1=0$. Thus an alternative G-formula approach for the hypothetical estimand is to only specify and fit a model among those patients who did not experience an ICE ($A_1=0$), i.e.\ for $E(Y|A_0=a_0,A_1=0,L_0,L_1)$. For example if we are interested in the potential outcome under a particular value of the treatment $A_0=a_0$ we might assume that $E(Y|A_{0}=a_{0},A_1=0,L_0,L_{1})=\beta_{0}^{a_0} + \beta_{1}^{a_0} L_{0} + \beta_{2}^{a_0} L_{1}$. This model can be fitted by ordinary least squares to those randomised to $A_{0}=a_{0}$ and for whom $A_{1}=0$, giving estimates $\hat{\beta}_{0}^{a_0}$, $\hat{\beta}_{1}^{a_0}$ and $\hat{\beta}_{2}^{a_0}$. The G-formula estimator is then equal to
\begin{equation}
     \widehat{E}(Y^{a_0,a_1=0}) = \frac{\sum^{n}_{i=1} I(A_{0,i}=a_0) \left\{ \hat{\beta}_{0}^{a_0} + \hat{\beta}_{1}^{a_0} L_{0,i} + \hat{\beta}_{2}^{a_0} L_{1,i} \right\}}{\sum^{n}_{i=1} I(A_{0,i}=a_0)}
     \label{gformulaLinMod}
\end{equation}
The approach which specifies and fits a model to only those patients who did not experience the ICE makes fewer assumptions than the one which specifies and fits a model to all observed data. The approach which uses all the data has the potential to give improved precision of estimates (and hence greater power), but is more vulnerable to model misspecification. Similarly, there is trade off between fitting a single model and including the randomised treatment $A_0$ as covariate for increased precision or fitting separate models by treatment arm to relax modelling assumptions. The different modelling alternatives can also be combined e.g. separate models by treatment arm among ICE-free. 

\subsubsection{Inverse probability of treatment weighting} \label{sec:IPW}
We now apply the general IPW estimator described in Section \ref{review_ipw}. Applied to the current setting, the IPW estimator of the mean of $Y^{a_0,a_1=0}$ from equation \eqref{eq:ipw_gen_timevarying} is given by
\begin{equation}
    \frac{\sum^{n}_{i=1} I(\bar{A}_{i}=(a_0,0)) W_{i} Y_{i}}{\sum^{n}_{i=1} I(\bar{A}_{i}=(a_0,0)) W_{i}}
    \label{eq:ipw_oneICE}
\end{equation}
This is a weighted mean of the outcomes from those patients who were randomised to treatment $a_0$ and in whom the ICE did not occur. Under 1:1 randomisation $P(A_0=a_0|L_0)=0.5$, and so the weights $W_i$ are given by
\begin{align}
W_{i} &= \frac{1}{0.5} \times \frac{1}{f(A_{1,i}|A_{0,i},L_{0,i},L_{1,i})} \nonumber \\
&=  \frac{2}{f(A_{1,i}|A_{0,i},L_{0,i},L_{1,i})} \label{eq:w_ipw_oneICE} \end{align}
Since the occurrence of the ICE is typically not under the investigator's full control, we must postulate and fit a model for $P(A_1=0|A_0,L_0,L_1)$. For example, we could fit a logistic regression for $A_1$ with $A_0$, $L_0$ and $L_1$ as covariates. Here the logistic models to estimate the weights could also be fitted separately by treatment arm. As with G-formula, there is a similar trade off of making the models less vulnerable to model misspecification but at the expense of precision. 

\subsection{Missing data approaches}

\subsubsection{Likelihood based missing data approaches}
As described in Section \ref{sec:intro}, currently a commonly adopted estimation approach when targeting hypothetical estimands with continuous endpoints is to fit a linear mixed model to the repeated measures of the outcome variable using maximum likelihood, after deleting any outcome measurements which were made after the ICE took place. These are based on assuming the resulting missing data are missing at random (MAR).

\begin{figure}[hbt!]
    \centering
    \begin{tikzpicture}
    
    \node (L_0) at (-1.00,2.00) {$L_0$};
    \node (A_0) at (0.00,0.00) {$A_0$};
    \node (L_1) at (2.00,2.00) {$L_1$};
    \node (A_1) at (3.00,0.00) {$A_1$};
    \node[draw,rectangle] (a_1) at (4.00,0.00) {$a_1=0$};
    \node (Y) at (7.00,0.00) {$Y^{a_1=0}$};
    
    \draw [->] (L_0) edge (A_1);
    \draw [->] (L_0) edge (L_1);
    \draw [->] (L_0) to[out=90,in=90] (Y);
    \draw [->] (A_0) edge (A_1);
    \draw [->] (A_0) edge (L_1);
    \draw [->] (A_0) to[out=-90,in=-90] (Y);
    \draw [->] (L_1) edge (A_1);
    \draw [->] (L_1) edge (Y);
    \draw [->] (a_1) edge (Y);
    \end{tikzpicture} 
    \caption{Single-world intervention graph (SWIG) resulting from the DAG shown in Figure \ref{fig:DAG_fixed}, intervening to set the ICE to $a_{1}=0$.}
    \label{fig:SWIG_fixed_mar}
\end{figure}
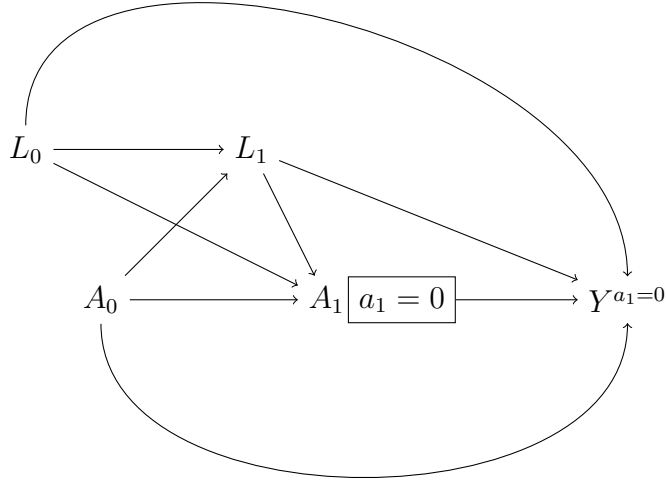

We will use the theory of DAGs to check whether MAR holds for the missing hypothetical outcomes under the DAG shown in Figure \ref{fig:DAG_fixed}. To do so, we make use of an extension of DAGs: single world intevention graphs (SWIGs), as described by \cite{Richardson2013SWIGs}. A SWIG takes as its input a DAG, and shows the graph that would result under an intervention which fixes the values of certain variables. Figure \ref{fig:SWIG_fixed_mar} shows the SWIG that results if we intervene to set the ICE $a_1=0$. Each variable intervened on is split into two parts, the first (in capitals) which denotes the original variable, taking whatever value it would naturally take (without intervention), and the second part (lower case) indicating the intervened value. Variables which are affected by  those variables intervened on are changed to their potential outcome value under the specified values of the intervened variables. Thus when we intervene to set $a_1=0$, $Y$ becomes $Y^{a_1=0}$.

Unlike the DAG, the SWIG in Figure \ref{fig:SWIG_fixed_mar} contains the partially observed potential outcomes $Y^{a_1=0}$ of interest under the hypothetical estimand. To check MAR, we note that the indicator of missingness in the hypothetical outcomes $Y^{a_{1}=0}$ is identical to the ICE variable $A_1$, since it is those individuals with $A_{1}=1$ for which $Y^{a_{1}=0}$ is missing. MAR here means that the missingness indicator $A_1$ is independent of the partially observed $Y^{a_{1}=0}$ conditional on $A_0$, $L_0$ and $L_{1}$. This conditional independence can be read off from the SWIG, since once we condition on $A_0$, $L_0$ and $L_{1}$ there are no open paths from $A_1$ to $Y^{a_{1}=0}$. Analogous to the sequential exchangeability assumption, we emphasize that we must ensure the variables used as $L_0$ and $L_1$ in estimation do include all common causes of the ICE $A_1$ and the outcome $Y$. For example, while $L_0$ and $L_1$ may typically need to include measurements of the outcome variable at baseline and the intermediate time point, there will generally be additional common causes of the ICE and outcome $Y$, and these must be included in $L_0$ and $L_1$ for the DAG in Figure \ref{fig:DAG_fixed} to be correct.

Since the potential outcomes $Y^{a_{1}=0}$ which are of interest for the hypothetical estimand are MAR given $A_0$, $L_0$ and $L_{1}$, it follows that an observed data likelihood analysis assuming the missingness is ignorable will give consistent estimates in this scenario under the previously stated assumptions and provided the full data model assumed is correctly specified. These conclusions are in agreement with those of \citet{holzhauer2015choice}, who proposed fitting a joint mixed effects model for $L_0,L_1,Y$
conditional on $A_0$, after deleting post ICE outcomes, or alternatively use of multiple imputation to impute $Y^{a_{1}=0}$ in those with $A_1=1$, again ensuring that $L_0,L_1,A_0$ are included in the imputation model.

We now show that particular observed data likelihood based estimators are identical to particular G-formula estimators. Consider the data on $L_{0}$, $L_{1}$ and $Y$ in treatment group $A_{0}=a_{0}$, after deleting any $Y$ values for individuals with $A_{1}=1$. Suppose that $L_1$ is a single variable and that we fit a bivariate normal model (a type of linear mixed model) to the resulting $(L_1,Y^{a_1=0})$ data in group $A_{0}=a_{0}$ assuming MAR, with the means of $L_{1}$ and $Y^{a_1=0}$ depending linearly on $L_0$ but with separate coefficients, and an unstructured covariance matrix. The bivariate normal model implies that
\begin{align*}
    E(Y^{a_1=0}|A_{0}=a_{0},L_{0},L_{1}) &=\beta^{a_0}_{20}+\beta^{a_0}_{21} L_{0} + \beta^{a_0}_{22} L_{1} \\
    E(L_1|A_{0}=a_{0},L_{0}) &=\beta_{10}^{a_0}+\beta_{11}^{a_0} L_{0} 
\end{align*}
The observed data likelihood function under MAR factorises (Section 7.2 of Little and Rubin 2019) such that the MLEs of the parameters in these two models are obtained by fitting the $Y^{a_1=0}$ model among those with $Y^{a_1=0}$ observed (here meaning $A_{1}=0$) and for the $L_1$ model using all patients. Then we have that
\begin{align*}
E(Y^{a_1=0} | A_{0}=a_{0},L_0)&= E\left\{E(Y^{a_1=0}|A_{0}=a_{0}, L_0, L_1) | A_{0}=a_{0} ,L_0 \right\} \\
&= E( \beta_{20}^{a_0} + \beta_{21}^{a_0} L_{0} + \beta_{22}^{a_0} L_{1} | A_{0}=a_{0}, L_0) \\
&= \beta_{20}^{a_0} + \beta_{21}^{a_0} L_{0} + \beta_{22}^{a_0} (\beta_{10}^{a_0}+\beta_{11}^{a_0} L_{0} )
\end{align*}
Taking expectations of this conditional on $A_0=a_0$ we have
\begin{align*}
E(Y^{a_1=0} | A_{0}=a_{0})&= \beta_{20}^{a_0} + \beta_{21}^{a_0} E(L_{0}|A_0=a_0) + \beta_{22}^{a_0} (\beta_{10}^{a_0}+\beta_{11}^{a_0} E(L_{0}|A_0=a_0) )
\end{align*}
Taking $\hat{E}(L_{0}|A_{0}=a_{0})=\frac{\sum^{n}_{i=1} I(A_{0,i}=a_0) L_{0,i}}{\sum^{n}_{i=1} I(A_{0,i}=a_0)}$ as the non-parametric MLE of $E(L_0|A_0=a_0)$, by the invariance property of MLE the MLE of $E(Y^{a_1=0}|A_0=a_0)$ is given by
\begin{align*}
    \hat{E}(Y^{a_1=0} | A_{0}=a_{0})&= \hat{\beta}_{20}^{a_0} + \hat{\beta}_{21}^{a_0} \hat{E}(L_{0}|A_0=a_0) + \hat{\beta}_{22}^{a_0} (\hat{\beta}_{10}^{a_0}+\hat{\beta}_{11}^{a_0} \hat{E}(L_{0}|A_0=a_0) ) 
\end{align*}
The model for $L_1$ is fitted to all those with $A_0=a_0$. A property of ordinary least squares estimators is that the sample mean of the dependent variable (here $L_{1}$) is equal to the predicted value of the dependent variable when the covariate is set to its sample mean, such that 
\begin{align*}
    \hat{\beta}_{10}^{a_0} + \hat{\beta}_{11}^{a_0} \hat{E}(L_{0}|A_0=a_0) =
    \frac{\sum^{n}_{i=1} I(A_{0,i}=a_0) L_{1,i}}{\sum^{n}_{i=1} I(A_{0,i}=a_0)}.
\end{align*}
It follows that 
\begin{align*}
    \hat{E}(Y^{a_1=0} | A_{0}=a_{0})&=
    \frac{\sum^{n}_{i=1} I(A_{0,i}=a_0) \left(\hat{\beta}_{20}^{a_0} + \hat{\beta}_{21}^{a_0} L_{0,i} + \hat{\beta}_{22}^{a_0} L_{1,i}  \right)}{\sum^{n}_{i=1} I(A_{0,i}=a_0)}
\end{align*}
which is identical to the G-formula estimator given in equation \eqref{gformulaLinMod}.

More commonly a linear mixed model is fitted which assumes a common covariance matrix for $(L_{1},Y^{a_{1}=0})$ across the two randomised groups with mean effects of $A_0$ and $L_0$. A similar argument to the one above shows that the resulting estimator is a G-formula estimator where we fit a single model for $Y^{a_1=0}$ to both randomised groups, with $A_0$ as a covariate (in addition to $L_0$ and $L_1$).

Lastly, we note that a multiple imputation (MI) analysis assuming MAR and based on the same modelling assumptions as the likelihood based analysis is (up to Monte-Carlo noise) equivalent to the likelihood based analysis. Thus corresponding multiple imputation estimates which delete data after the ICE occurs are equivalent to particular G-formula implementations. The equivalence of imputation approaches and G-formula was previously discussed by \citet{Westreich2015Imputation}.

\subsubsection{Inverse probability of missing weighting}
The potential outcomes of interest among those randomised to $A_0=a_0$ are $Y^{a_1=0}$. Since the event that $A_1=0$ is precisely the indicator of observation of the potential outcome of interest, the standard IPW missing data estimator \citep{seaman2013review} for $E(Y^{a_1=0}|A_0=a_0)$ can be seen to be identical to the `causal inference' IPW estimator given in equation \eqref{eq:ipw_oneICE}.

\subsection{Deterministic intercurrent events}\label{sec:deterministic}
In some trials the intercurrent event could be discontinuation of randomised treatment or addition of rescue treatment. In some therapy areas, e.g. diabetes, the protocol specifies that rescue treatment be given at/following a visit at time $k$ if and only if a measurement of glycemic control (e.g. via FPG or HbA1c) exceeds some threshold. This has been termed a deterministic MAR situation \citep{holzhauer2015choice}, with missing data estimation approaches advocated.

In such situations the positivity assumption is violated if the protocol was followed. This implies that the hypothetical estimand cannot be nonparametrically identified, which essentially means it cannot be estimated without making untestable modelling assumptions. Parametric G-formula, likelihood and MI approaches can provide consistent estimates, but only by extrapolating beyond the data. In particular, they must predict the no ICE potential outcomes for those individuals who did in fact have an ICE. When intercurrent events occur deterministically, there are no similar patients in terms of $A_0,L_0,L_1$ who did not have the ICE and hence have $Y^{a_1=0}$ observed. If the extrapolation implied by the model is correct, we can obtain consistent estimates. However, from the data alone, we have no basis on which to judge whether the extrapolation is justified. In such cases, the extrapolation can arguably only be justified on the basis of external evidence, since the data offer no information about whether the extrapolation is reliable.

In contrast, if the IPW approach is used, and the model for ICE/missingness correctly incorporates the deterministic ICE mechanism, for those with $A_{1}=0$ because $P(A_{1}=0|A_0,L_0,L_1)=1$ their true weight will be 2 as per equation \eqref{eq:w_ipw_oneICE}, as well as for those with $A_{1}=1$ because $P(A_{1}=1|A_0,L_0,L_1)=1$. In this case, the estimator in equation \ref{eq:ipw_oneICE} would simply be the unweighted average of outcomes in those with $A_1=0$, which in general will give a biased estimate.


\section{Intercurrent events at multiple time points}
\label{sec:multipletime}
\begin{figure}[hbt!]
    \centering
    \begin{tikzpicture}
    \node (L_0) at (-1.00,2.00) {$L_0$};
    \node (A_0) at (0.00,0.00) {$A_0$};
    \node (A_1) at (3.00,0.00) {$A_1$};
    \node (A_2) at (6.00,0.00) {$A_2$};
    \node (L_1) at (2.00,2.00) {$L_1$};
    \node (L_2) at (5.00,2.00) {$L_2$};
    \node (Y) at (9.00,0.00) {$Y$};
    
    \draw [->] (L_0) edge (L_1);
    \draw [->] (L_0) edge (A_1);
    \draw [->] (L_0) to[out=90,in=90] (L_2);
    \draw [->] (L_0) edge (A_2);
    \draw [->] (L_0) to[out=90,in=90] (Y);
    
    \draw [->] (A_0) edge (A_1);
    \draw [->] (A_0) to[out=-90,in=-90] (A_2);
    \draw [->] (A_0) to[out=-90,in=-90] (Y);
    \draw [->] (A_0) edge (L_1);
    \draw [->] (A_0) to[out=90,in=90] (L_2);
    
    \draw [->] (A_1) edge (A_2);
    \draw [->] (A_1) to[out=-90,in=-90] (Y);
    \draw [->] (A_1) edge (L_2);
    
    \draw [->] (A_2) edge (Y);
    
    \draw [->] (L_1) edge (L_2);
    \draw [->] (L_1) edge (A_1);
    \draw [->] (L_1) edge (A_2);
    \draw [->] (L_1) edge (Y);
    
    \draw [->] (L_2) edge (A_2);
    \draw [->] (L_2) edge (Y);
    
    \end{tikzpicture} 
    
    \caption{Directed acyclic graph (DAG) representation of randomised trial with ICE occurring at two time points.}
    \label{fig:DAG_twotimes}
\end{figure}
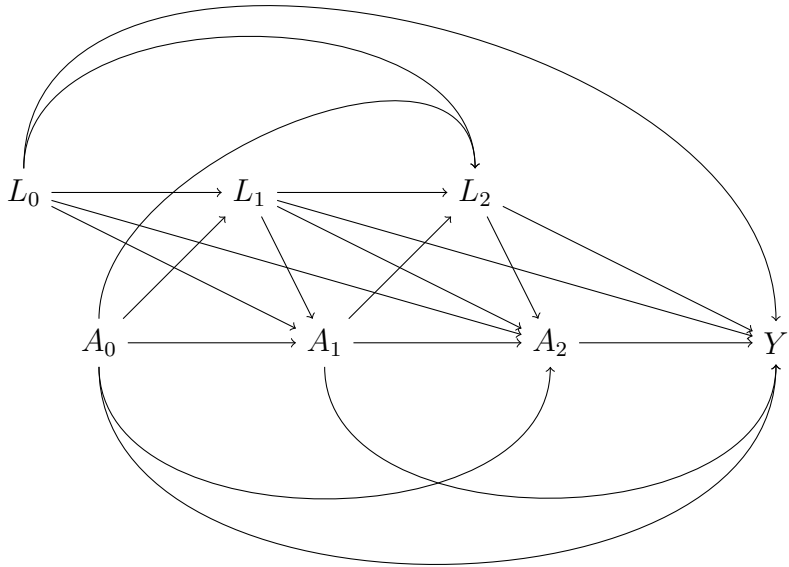

We now consider the more realistic setting where the ICE can occur at multiple time points. We consider the case where the ICE can occur at two time points, denoted $A_1$ and $A_2$, and note that our conclusions in this situation can be easily extended to the general setup with more time points. The hypothetical potential outcomes of interest now are $Y^{a_0,a_1=0,a_2=0}$ for $a_0=0$ and $a_0=1$. Figure \ref{fig:DAG_twotimes} shows the DAG for this situation.

\subsection{Causal inference approaches}

\subsubsection{G-formula}
Using equation \eqref{eq:gform_timevar_exp} we have that
\begin{align*}
    E(Y^{\bar{a}}) &=  \int_{l_0} \int_{l_1} \int_{l_2} E(Y|\bar{a},\bar{l})  f(l_2|\bar{a}_1,\bar{l}_1)f(l_1|a_0,l_0) f(l_0) dl_2 dl_1 dl_0
\end{align*}
As before because of randomisation $f(l_0)=f(l_0|a_0)$ and so we can write
\begin{align}
    E(Y^{\bar{a}}) &=  \int_{l_0} \int_{l_1} \int_{l_2} E(Y|\bar{a},\bar{l})  f(l_2|\bar{a}_{1},\bar{l}_1)f(l_1,l_0|a_0)  dl_2 dl_1 dl_0  \label{gFormulaMultTimeEmpiricalL0L1}
\end{align}
To construct an estimator based on this, we specify and fit a model for $E(Y|\bar{A},\bar{L})$ and for $f(L_2|\bar{A}_{1},L_0,L_1)$, while for $f(L_1,L_0|A_0=a_0)$ we can again avoid modelling it by using the empirical distribution of $(L_1,L_0)$ in those randomised to treatment $a_0$. As before, we could fit models for $E(Y|\bar{A},\bar{L})$ and $f(L_2|\bar{A}_{1},L_0,L_1)$ using data from all patients with suitable covariate effect specification, or instead restrict them to those whose treatment history $\bar{A}$ equals the hypothetical estimand regime of interest $\bar{a}=(a_0,0,0)$. The former may be more efficient, but it requires specification of more modelling assumptions. Taking the latter approach, suppose we assume that 
\begin{align}
    E(Y |A_0=a_{0},A_1=0,A_2=0,L_0, L_1,L_2) &= \beta_{30}^{a_0} + \beta_{31}^{a_0} L_{0} + \beta_{32}^{a_0} L_{1} + \beta_{33}^{a_0} L_2 \nonumber \\
    E(L_2 | A_0=a_0, A_1=0, L_0, L_1) &= \beta_{20}^{a_0} + \beta_{21}^{a_0} L_{0} + \beta_{22}^{a_0} L_{1}
    \label{conditionMeansTwoTimes}
\end{align}
Then we have that
\begin{align*}
    & E\left\{ E(Y |A_0=a_{0},A_1=0,A_2=0,L_0, L_1,L_2) | A_0=a_0,A_1=0,L_0,L_1\right\} \\
    &= E\left\{ \beta_{30}^{a_0} + \beta_{31}^{a_0} L_{0} + \beta_{32}^{a_0} L_{1} + \beta_{33}^{a_0} L_2| A_0=a_0,A_1=0,L_0,L_1\right\} \\
    &= \beta_{30}^{a_0} + \beta_{31}^{a_0} L_0 + \beta_{32}^{a_0} L_1 + \beta_{33}^{a_0} E(L_2|A_0=a_0,A_1=0,L_0,L_1) \\
    &= \beta_{30}^{a_0} + \beta_{31}^{a_0} L_0 + \beta_{32}^{a_0} L_1 + \beta_{33}^{a_0} ( \beta_{20}^{a_0} + \beta_{21}^{a_0} L_{0} + \beta_{22}^{a_0} L_{1}) \\
    &= \beta_{30}^{a_0} + \beta_{33}^{a_0}\beta_{20}^{a_0} + (\beta_{31}^{a_0}+\beta_{33}^{a_0}\beta_{21}^{a_0}) L_0 + (\beta_{32}^{a_0}+\beta_{33}^{a_0}\beta_{22}^{a_0}) L_1
\end{align*}
Then using equation \eqref{gFormulaMultTimeEmpiricalL0L1} 
our G-formula estimator is
\begin{align}
    \frac{\sum^{n}_{i=1} I(A_{0,i}=a_0) \{\hat{\beta}_{30}^{a_0} + \hat{\beta}_{33}^{a_0}\hat{\beta}_{20}^{a_0} + (\hat{\beta}_{31}^{a_0} + \hat{\beta}_{33}^{a_0} \hat{\beta}_{21}^{a_0}) L_{0,i} + (\hat{\beta}_{32}^{a_0} + \hat{\beta}_{33}^{a_0} \hat{\beta}_{22}^{a_0}) L_{1,i} \} }{\sum^{n}_{i=1} I(A_{0,i}=a_0)}
    \label{eq:gform_multtime_linear}
\end{align}
In words, for each patient randomised to treatment $a_0$, this G-formula estimator first predicts $L^{a_0,a_1=0}_2$ under the hypothetical no ICE scenario and then predicts $Y^{a_0,a_1=0,a_2=0}$. Finally,  it averages these predictions across the patients randomised to $a_0$.

\subsubsection{Inverse probability of treatment weighting} \label{sec:ipw_twoICE}

For this setting when the ICE can occur at two time points, the IPW estimator for the mean of $Y^{a_0,a_1=0,a_2=0}$ is a weighted mean of the outcomes from those patients who were randomised to treatment $a_0$ and in whom the ICE did not occur at either of the two intermediate time points. The weights are as defined in equation \eqref{eq:ipwGeneralUnstabilised}. 

Unlike in the setting considered in Section \ref{sec:singletime}, we could now choose to model the occurrence of ICE at each time point using all patients, with earlier occurrence of ICE as a covariate. However, since in the end we only need weights for those patients who did not experience the ICE at either time point, we might choose instead to model the occurrence of the ICE at time $k$ only among those who had not up to time $k$ experienced an ICE. Thus like the G-formula, one has some flexibility and choice about which data to use and what modelling assumptions to make. Indeed, one  may wish to avoid modelling how the occurrence of an ICE depends on the past among those who have already experienced an ICE.

\subsection{Missing data approaches}
\subsubsection{Likelihood based missing data approaches}
As noted earlier, missing data methods have previously been advocated to and applied for estimating hypothetical estimands by fitting mixed models to the repeated measurements of outcomes after excluding any post ICE outcomes. This leads to a so called monotone missingness pattern. We now show using a SWIG derived from the DAG in Figure \ref{fig:DAG_twotimes} that the hypothetical potential outcomes of interest are again MAR.

The full data under the hypothetical estimand is now $(L_1,L_2^{a_1=0},Y^{a_1=0,a_2=0})$. There are missing values in $L_2^{a_1=0}$ and $Y^{a_1=0,a_2=0}$, and if $L_2^{a_1=0}$ is missing for an individual then so is $Y^{a_1=0,a_2=0}$. This is analogous to monotone dropout in a longitudinal study. In this context, MAR can be expressed as saying that at any given time, among those subjects who have not yet dropped out, the probability of dropout before the next follow-up visit is independent of future outcomes given the past outcomes  \citep{daniels2008missing}.

\begin{figure}[hbt!]
    \centering
    \begin{tikzpicture}
    \node (L_0) at (-1.00,2.00) {$L_0$};
    \node (A_0) at (0.00,0.00) {$A_0$};
    \node (L_1) at (2.00,2.00) {$L_1$};
    \node (L_2) at (6.00,2.00) {$L_2^{a_1=0}$};
    \node (Y) at (13.00,0.00) {$Y^{a_1=0,a_2=0}$};
    \node (A_1) at (3.00,0.00) {$A_1$};
    \node [draw,rectangle] (a_1) at (4.20,0.00) {$a_1=0$};
    \node (A_2) at (8.00,0.00) {$A_2^{a_1=0}$};
    \node [draw,rectangle] (a_2) at (9.40,0.00) {$a_2=0$};
    
    \draw [->] (L_0) edge (L_1);
    \draw [->] (L_0) edge (A_1);
    \draw [->] (L_0) to[out=90,in=90] (L_2);
    \draw [->] (L_0) edge (A_2);
    \draw [->] (L_0) to[out=90,in=90] (Y);
    
    \draw [->] (A_0) edge (A_1);
    \draw [->] (A_0) to[out=-90,in=-90] (A_2);
    \draw [->] (A_0) to[out=-90,in=-90] (Y);
    
    \draw [->] (A_0) edge (L_1);
    \draw [->] (A_0) to[out=90,in=90] (L_2);
    
    \draw [->] (a_1) edge (A_2);
    \draw [->] (a_1) to[out=-90,in=-90] (Y);
    \draw [->] (a_1) edge (L_2);
    
    \draw [->] (a_2) edge (Y);
    
    \draw [->] (L_1) edge (L_2);
    \draw [->] (L_1) edge (A_1);
    \draw [->] (L_1) edge (A_2);
    \draw [->] (L_1) edge (Y);
    
    \draw [->] (L_2) edge (A_2);
    \draw [->] (L_2) edge (Y);
    
    \end{tikzpicture} 
    \caption{SWIG resulting from DAG shown in Figure \ref{fig:DAG_twotimes}, intervening to set $a_1=0$ and $a_2=0$.}
    \label{fig:SWIG_twotimes}
\end{figure}
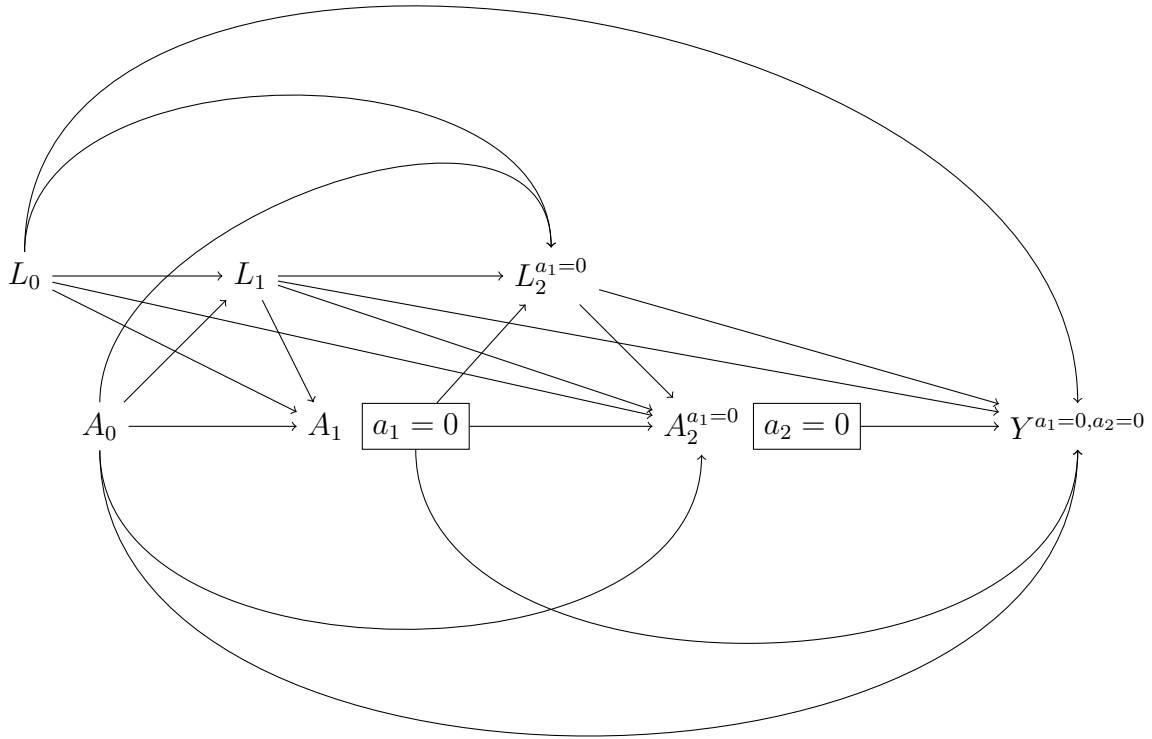

Figure \ref{fig:SWIG_twotimes} shows the SWIG resulting from the DAG in Figure \ref{fig:DAG_twotimes} if we intervene to set $a_1=0$ and $a_2=0$, and we can use this to check MAR is satisfied. First we can immediately confirm from the SWIG that $A_1 \indep (L_2^{a_1=0},Y^{a_1=0,a_2=0}) | A_0,L_1,L_0$. Next we must check that $A_2 \indep Y^{a_1=0,a_2=0} | A_1=0, A_0, L_2, L_1, L_0$. For this, note that in those with $A_1=0$, by the consistency assumption $L_2=L_2^{a_1=0}$ and $A_2=A_2^{a_1=0}$, and so this assumption is equivalent to $A_2^{a_1=0} \indep Y^{a_1=0,a_2=0} | A_1=0, A_0, L_2^{a_1=0}, L_1, L_0 $, and the SWIG shows that this conditional independence condition indeed holds. We emphasize again, that if there are, as there typically would be, common causes of ICE occurrence and final outcome additional to the repeated measurements of outcome, they must be included in $L_0$, $L_1$, $L_2$. Others have previously discussed the use of missing data methods assuming MAR whereby data after the ICE occurs are excluded from the analysis \citep{holzhauer2015choice,mallinckrodt2020aligning}. By using the machinery of causal diagrams, we are able to clarify the conditions under which the MAR assumption would be satisfied - namely that we have measured and conditioned on all common causes of the ICE variables (here $A_1$ and $A_2$) and the final outcome of interest (here $Y$).

As was the case with an ICE occurring at a single time point, in the Supplementary Material we show that particular `missing data estimators' and particular `causal inference estimators' are equivalent. The fact that missing data approaches are valid and are identical to particular causal inference estimators in this setting is perhaps at first sight surprising, since it is often said that a setting with time-dependent confounding affected by treatment is one in which `standard approaches' are invalid, with more complex g-methods required instead. For further details, see the Supplementary Material. 


\subsubsection{Inverse probability of missing weighting}

As before we note that we are considering the ICE as a `treatment' and it is also the missing indicator because we are only interested in the potential outcomes $Y^{a_1=0,a_2=0}$ which would occur in the absence of the ICE. However, when the ICE can occur at more than one time point, the implementation of IPW for missingness is equivalent to the one described in Section \ref{sec:ipw_twoICE} where we fit the models to estimate the weights at time point $k$ restricted to those with $\overline{A}_{1:k-1}=\bar{0}$. 

\section{Simulations}
\label{sec:sims}

To illustrate and compare the different estimation methods' performance, we conducted simulations with probabilistic or deterministic ICE occurrence. We also explore the impacts of misspecification in the models used by the different estimation approaches. In what follows, $\bar{c}$ denotes a column vector of suitable length for conformability with all elements equal to the scalar $c$.



\subsection{Probabilistic intercurrent event}

We generated $10,000$ datasets of $500$ subjects where the treatment $A_0$ was assigned 1:1 at random and the ICE could occur at 5 different time points during follow-up ($K=5$) as follows: 

\begin{itemize}
    \item $A_0 \sim Ber(0.5)$
    \item $L_0 \sim \mathcal{N}(0,1)$
    \item $L_k \sim \mathcal{N}(\overline{0.3}^T \bar{L}_{k-1} + \overline{0.2}^T \bar{A}_{k-1},1)$ for $k = 1 : 5$
    \item $A_k \sim Ber \Big( \text{expit}(-3+ \overline{0.2}^T \bar{L}_k + \overline{0.4}^T \bar{A}_{k-1}) \Big)$ for $k = 1 : 5$
    \item $Y \sim \mathcal{N}(\overline{0.2}^T \bar{L} + 0.5 A_0 + \overline{0.3}^T \bar{A}_{1:5},1) $
\end{itemize}
The parameter values were chosen so that on average 60-70 \% would be ICE free at the end of follow-up, for the different scenarios considered. 

Each of the generated datasets was analysed using the following methods:

\begin{enumerate}
    \item Naive: treatment effect estimated as the difference in mean outcome between the randomised treatment arms, among those who did not experience the ICE:
    \begin{align*}
        & \widehat{E}  (Y|A_0=1,\bar{A}_{1:5}=\bar{0})- \widehat{E}(Y|A_0=0,\bar{A}_{1:5}=\bar{0}) 
        \\ & = \frac{ \sum^n_{i=1} I(A_{i}=(1,\bar{0})) Y_i }{\sum^n_{i=1}  I(A_{i}=(1,\bar{0}))} - \frac{ \sum^n_{i=1}   I(\bar{A}_{i}=(0,\bar{0})) Y_i}{\sum^n_{i=1}  I(\bar{A}_{i}=(0,\bar{0}))}
    \end{align*}
    
    \item G-formula using all data: 
    \begin{enumerate}
        \item First fit linear models for $L$ at each time point $k \in 1:5$, given the observed covariate history until time $k$ $(\bar{A}_{k-1}, \bar{L}_{k-1})$, including everyone, regardless of whether they had the ICE at anytime during follow-up or not: $\widehat{E}(L_k|\bar{A}_{k-1},\bar{L}_{k-1}) $, with main effects of each past $A$ variable and each past $L$ variable.
        \item Linear model for $\widehat{E}(Y|\bar{A},\bar{L})$ is fitted including everyone, with main effects of each $A$ variable and each $L$ variable. 
        \item For every individual, a value of $\widehat{L}_{k,i}$ is predicted from the model at each time point, given their randomised treatment $A_{0,i}$, observed baseline covariate $L_{0,i}$, predicted covariate history up until that time point $k$ $(\widehat{\bar{L}}_{k-1,i})$, and setting $\bar{A}_{1:k}=0$. 
        \item For every individual, a value of $\widehat{Y}_i$ is predicted from the outcome model, given their randomised treatment $A_{0,i}$, observed baseline covariate $L_{0,i}$, full predicted covariate history $\widehat{\bar{L}}$ , and setting $\bar{A}_{1:5}=0$.
        \item The treatment effect estimate is the difference in the means of $\widehat{Y}_i$ between the $A_0=1$ and $A_0=0$ groups, i.e. $\hat{E}(\widehat{Y} | A_0=1) - \hat{E}(\widehat{Y} | A_0=0)$.
    \end{enumerate} 
    
    \item G-formula among ICE-free: similar to the previous one but the linear models for $L_k$s and $Y$ are fitted only including subjects ICE-free up until the corresponding time point.
    
    \item G-formula separately by treatment arm: similar to the first G-formula implementation but the linear models for $L$s and $Y$ are fitted separately by randomised treatment groups.
    
    \item G-formula among ICE-free separately by treatment arm: similar to the previous one where the linear models for $L$s and $Y$ are fitted separately by randomised treatment groups but among those ICE-free up until the corresponding time point.
    
    \item Inverse probability of ICE weighting:
    \begin{enumerate}
        \item First a logistic regression is fitted to estimate the probability of the occurrence of the ICE at each time point, given the randomised treatment, observed ICE history, and observed covariate history: $P(A_k=1| \bar{A}_{k-1}, \bar{L}_k)$ with main effects of each past $A$ variable and each past $L$ variable.
        \item The weights $W_i$ are calculated according to equation \eqref{eq:ipwGeneralUnstabilised}.
        \item The treatment effect estimate is calculated using equation \eqref{eq:ipw_gen_timevarying}. 
    \end{enumerate}
    
    \item Inverse probability of ICE weighting among ICE-free: this is similar to the previous method but the difference is that the logistic regression at each time point $k$ is fitted including only those ICE free until that time point ($\bar{A}_{1:k-1}=\bar{0}$), and there is no adjustment for past $A$s (which are all zero in this subset).
    
    \item Separate inverse probability of ICE weighting per arm treatment: similar to the first IPW version but fitting weight models separately by treatment arm. 
    
    \item Separate inverse probability of ICE weighting per arm treatment among ICE-free: similar to the previous version where the weight models are fitted separately by treatment arm but among those ICE-free up until the corresponding time point. 
    
    \item Multiple imputation with treatment as covariate using the {\tt mice} package:
    \begin{enumerate}
        \item First, all the observations ($L$s and $Y$) that occur after the first ICE are deleted.
        \item Sequential imputation of each $L$ is performed using normal linear regression, adjusting for main effects of $A_0$ and past $L$s.
        
        \item Imputation of $Y$ using normal linear regression, adjusting for $A_0$ and past $L$s.
        \item The treatment effect estimate was the difference in mean of $Y$ between $A_0=1$ and $A_0=0$ groups across the 10 imputed datasets. 
    \end{enumerate}
    
    \item Multiple imputation separately by treatment arm: similar to the previous one but the imputation models are constructed separately for each treatment arm. 
\end{enumerate} 

The results of the simulations are presented as box-and-whisker plots. For convenience, the extreme values were removed. Thus, any values larger than 1.5 times the interquartile range below or above the 25th percentile (lower hinge) and 75th percentile (upper hinge) respectively are not shown. The R code used for this simulations together with the plots containing the extreme values are freely available in the GitHub repository (URL: \url{https://github.com/colartep/hypothetical_estimands}). 

\begin{figure}
    \centering
    \includegraphics[width=\textwidth]{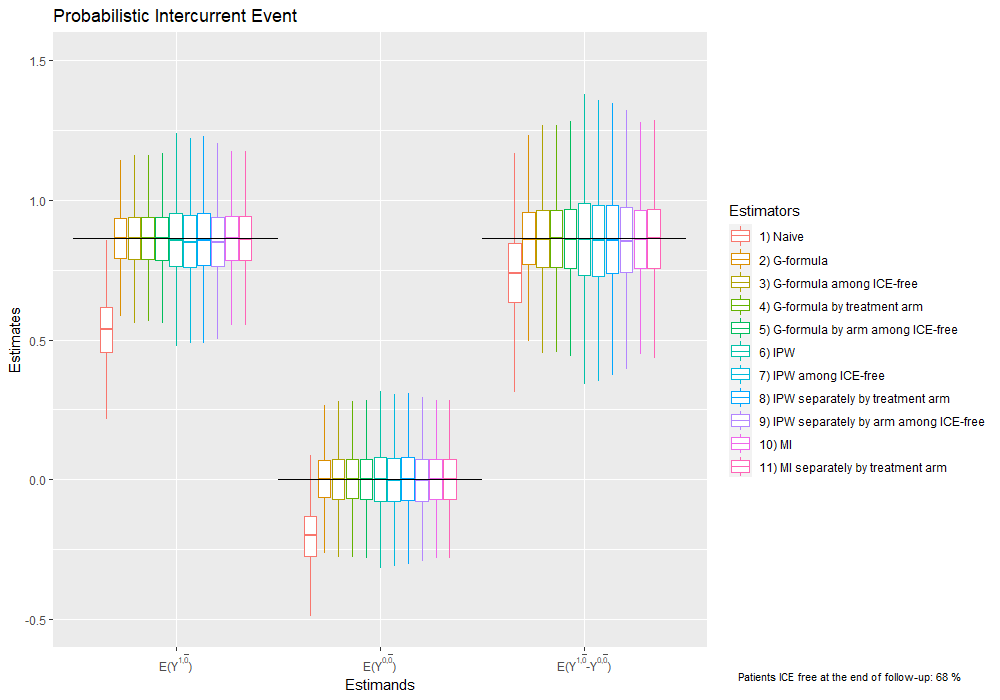}
    \caption{Simulation results showing estimates of potential outcome means under no ICE for $A_0=0$ and $A_0=1$ and their contrast, by different methods.}
    \label{fig:res_prob_ICE}
\end{figure}

Figure \ref{fig:res_prob_ICE} shows the comparison of the different estimators to estimate the mean outcome under treatment, control and their difference i.e. treatment effect, under no ICE. The naive estimator was biased for the potential outcome means and their contrast, as expected.

Since all models used were correctly specified, the different implementations of the G-formula, IPW and multiple imputation provided unbiased estimates. From the different G-formula versions, it is evident that when using all the available information the precision is improved, as expected. As anticipated, the IPW estimators were more variable than the G-formula estimators. For G-formula, IPW and MI, fitting separate models by randomised arm led to more variability in estimates, as one would expect.

The MI estimates were somewhat more variable than their G-formula counterparts, which is attributable to Monte-Carlo noise in the imputation process. If desired, this can be reduced by using a larger number of imputations.

\subsection{Deterministic switching}
The initial data generating mechanism was modified to generate an ICE which will only occur if the value of the time-varying covariate in that visit exceeds a threshold as follows:
\begin{equation}
  A_k =
    \begin{cases}
      1 & \text{if} \hspace{0.2cm} L_k \geq 1.5\\
      0 & \text{otherwise}\\
    \end{cases}
    \hspace{0.2cm} \text{for} \hspace{0.2cm}  k = 1 : 5
\end{equation}

\begin{figure}
    \centering
    \includegraphics[width=\textwidth]{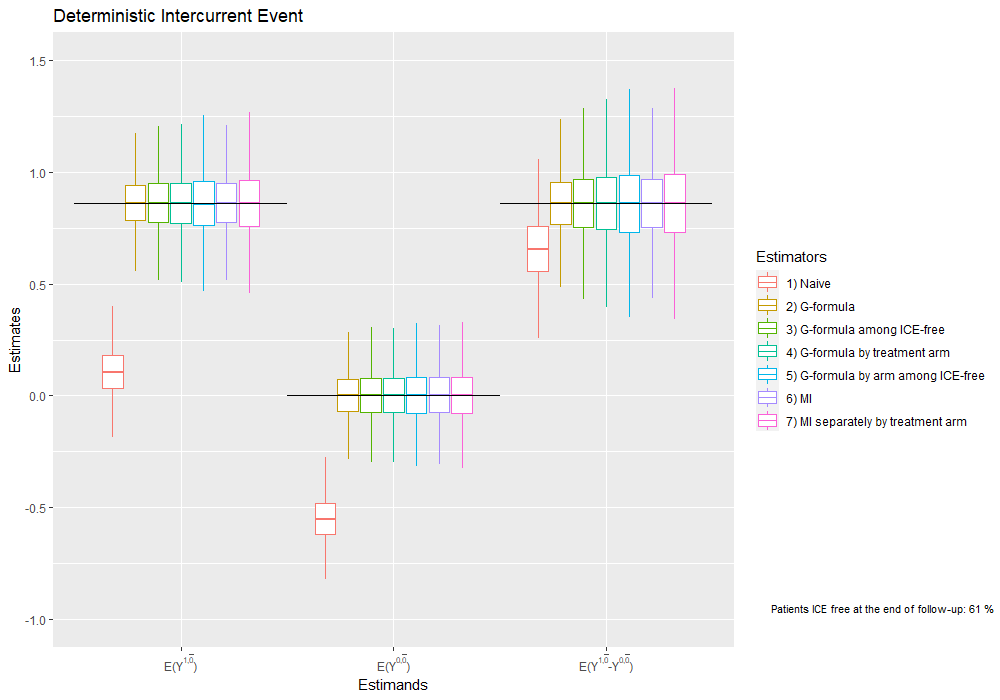}
    \caption{Simulation results showing the treatment effect as estimated by different methods, when the intercurrent event is determined non-randomly by the value of the time-varying covariates $L$.}
    \label{fig:res_deter_ICE}
\end{figure}

Figure \ref{fig:res_deter_ICE} shows the results under this deterministic setting. As described in Section \ref{sec:deterministic}, when the ICE is deterministic given the covariates, the weights are the same for everyone and the IPW is equivalent to an unweighted average among those ICE free which corresponds to the naive estimator here. Thus, we do not include the IPW estimators for this setting. 
The G-formula and MI estimators give unbiased results because the extrapolation beyond the observed data they make is based on correctly specified models. 

\subsection{Model misspecification}
To investigate the impacts of model misspecification, we introduce an additional term, namely ${L_0}^2$ to the $Y$ model in the data generating process. We then assess the impact on the estimation of the treatment effect  without adapting the estimators previously described. We also introduced the quadratic term to the $L_K$ model and $A_K$ model in turn to assess their respective impact. Finally we considered a setting where all the models had this additional term. 

Of note, if the direction of the bias is the same when estimating the two treatment regimens of interest, the bias may approximately cancel out for the treatment effect estimate. Thus, we allowed for differential magnitude and direction of the bias between treatment arms as follows:

\begin{itemize}
    \item $L_5 \sim \mathcal{N}(\overline{0.3}^T \bar{L}_{4} + \overline{0.2}^T \bar{A}_{4} + 2 {L_0}^2 A_0 - 0.5 {L_0}^2 (1-A_0),1)$ for $k = 1 : 5$
    \item $A_5 \sim Ber \Big( \text{expit}(-3+ \overline{0.2}^T \bar{L}_k + \overline{0.4}^T \bar{A}_{k-1} + 2 {L_0}^2 A_0 - 0.5 {L_0}^2 (1-A_0)) \Big)$ for $k = 1 : 5$
    \item $Y \sim \mathcal{N}(\overline{0.2}^T \bar{L} + 0.5 A_0 + \overline{0.3}^T \bar{A}_{1:5} + 2 {L_0}^2 A_0 - 0.5 {L_0}^2 (1-A_0),1) $
\end{itemize}

\begin{figure}
    \centering
    \includegraphics[width=\textwidth]{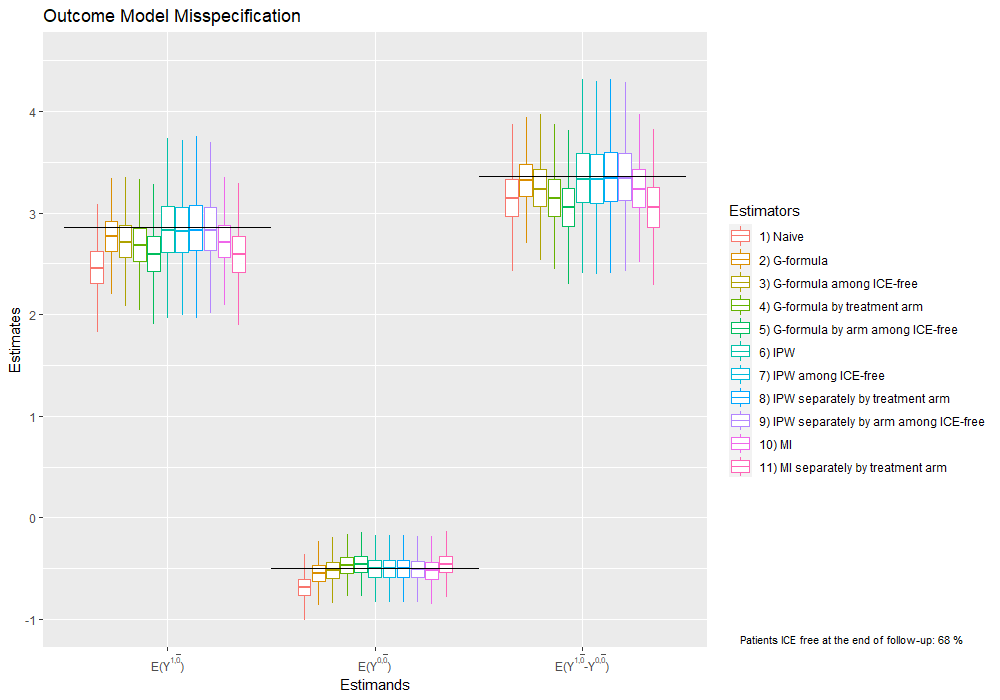}
    \caption{Simulation results showing the treatment effect as estimated by different methods, when the outcome model is misspecified. 
    }
    \label{fig:res_outcome_model_miss}
\end{figure}

\begin{figure}
    \centering
    \includegraphics[width=\textwidth]{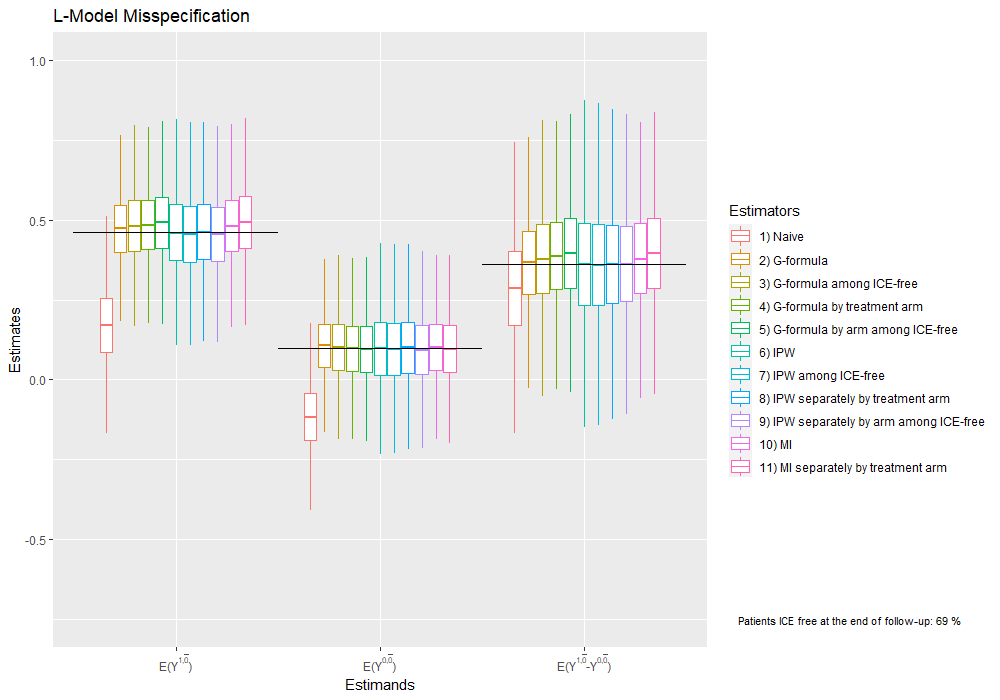}
    \caption{Simulation results showing the treatment effect as estimated by different methods, when the L-model is misspecified}
    \label{fig:res_L_model_miss}
\end{figure}

Figure \ref{fig:res_outcome_model_miss} presents the results when the outcome model is misspecified. Here the different implementations of the G-formula and multiple imputation give biased results as expected. As IPW does not rely on modelling the outcome, it is robust in this situation and provides unbiased results, although it is dramatically more variable. A similar situation is observed when the L-model is misspecified as shown in Figure \ref{fig:res_L_model_miss}.

\begin{figure}
    \centering
    \includegraphics[width=\textwidth]{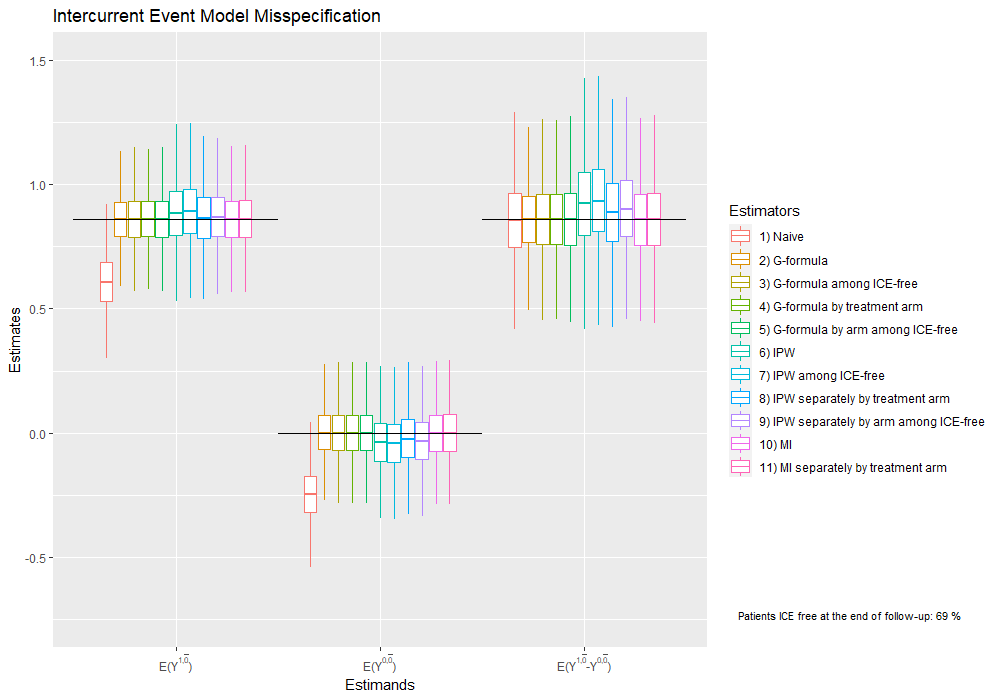}
    \caption{Simulation results showing the treatment effect as estimated by different methods, when the ICE model is misspecified}
    \label{fig:res_ICE_model_miss}
\end{figure}

The opposite situation is encountered when the ICE models are misspecified, as shown in Figure \ref{fig:res_ICE_model_miss}. Here, the IPW estimators give biased results, while G-formula and MI are not because the latter do not depend on modelling the ICE occurrence. When all models are misspecified, the different estimators considered fail to give unbiased estimates. 

\begin{figure}
    \centering
    \includegraphics[width=\textwidth]{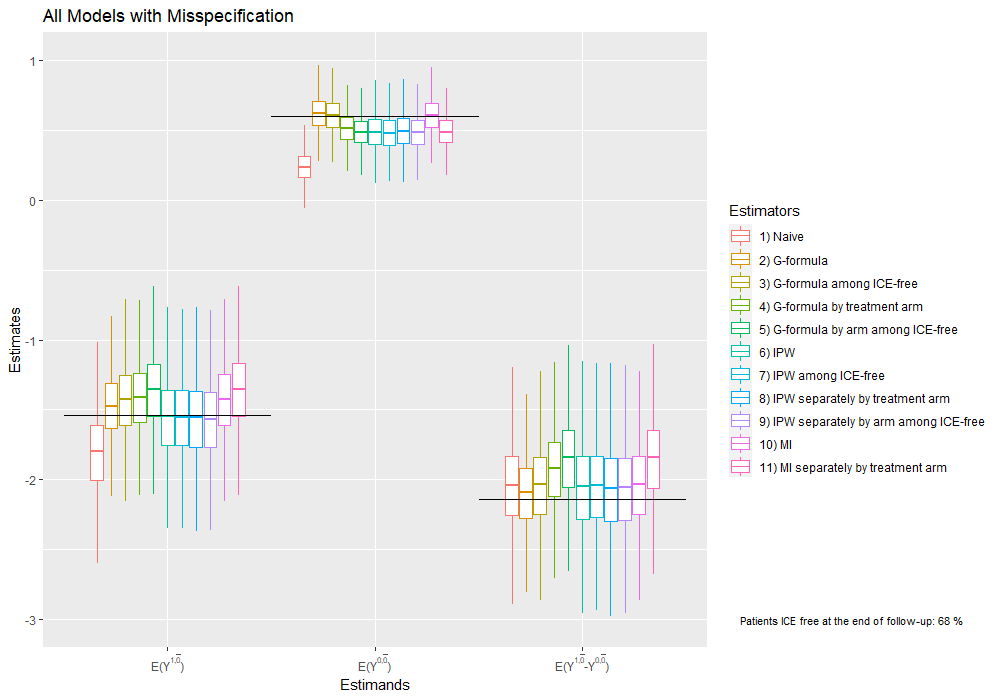}
    \caption{Simulation results showing the treatment effect as estimated by different methods, when the L, ICE and outcome models are all misspecified}
    \label{fig:res_all_models_miss}
\end{figure}

\section{Conclusion}
\label{sec:conc}

To target a hypothetical estimand under no ICE, we have considered different estimators arising from the causal inference and missing data literature. In doing so, we have shown that there are close connections between estimators in these groups. Indeed, we have shown that `missing data' likelihood based estimators (observed data likelihood or multiple imputation) applied after data post-ICE are deleted are implementations of the G-formula method from causal inference. Similarly, inverse probability of missing estimators for hypothetical estimands are also inverse probability of treatment estimators from causal inference.

We believe this unification is helpful to those analysing clinical trial data not least because the assumptions required for estimation, expressed via causal inference language, are arguably more easily understood than missing data assumptions. In particular, we believe DAGs can be very useful tools for graphically encoding what we may view as plausible for the causal relationships between variables, and from this, the validity of the sequential exchangeability assumption can be assessed. 

The causal inference lens also brings to the fore the importance of the positivity assumption. In trials where ICE occurrence is a deterministic function of biomarker values, such that the assumption is violated, estimation via likelihood methods or G-formula relies on extrapolation. The reasonableness of this extrapolation should be assessed on a case by case basis, taking into account the type of ICE, the disease context, and the extent of the extrapolation being made. Sensitivity analyses may be required given that the reliability of the extrapolation cannot be assessed from the observed data. Alternatively, in such cases one may choose to target a different estimand \citep{Michiels2021Balanced}.

It is worth noting that, in a given context, different hypothetical estimands can be defined for a particular ICE. For instance, in the case of rescue medication, we could conceive a hypothetical scenario where rescue medication was not available for the early stage of the trial i.e. no rescue medication in the first 6 months in a 36 months follow-up trial. The alternatives could be to set a shorter/longer period or to consider not having rescue medication at all. The hypothetical strategy could be regarded even more broadly and instead of considering an intervention to set the ICE to 0 for everyone, this could be set to a different value. For example, this could mean assessing the randomised treatment vs control treatment where everyone was to receive rescue medication. 

While we have shown that commonly used `missing data estimators' for hypothetical estimands correspond to certain causal inference estimators, we have also seen that there are additional implementations of G-formula and IPW which could be used instead. For example, contrary to current common practice, it is possible to use the full data, including intermediate and final outcomes assessed after the occurrence of the ICE, so long as suitable adjustment is made for past ICE occurrence. As seen in the simulations, these offer the potential for more precise estimates, at the expense of relying on more modelling assumptions.

There can be settings where it is plausible to borrow information from patients with ICEs. Consider a diabetes trial assessing the impact of a novel treatment with standard of care on achieving normal HbA1c levels. If a patient discontinues the novel treatment and still achieves normal HbA1c levels, they would probably have also had a positive outcome had they continued on treatment. 



It is worth noting that the hypothetical estimand is particularly relevant to deal with ICEs that can be intervened on. For instance we could conceive a trial where we could intervene to avoid treatment interruptions due to administrative reasons such as the ones derived from government enforced closures during the current COVID-19 pandemic. In contrast, it would probably be less sensible to consider a world were we could intervene to avoid treatment discontinuation due to adverse effects. For these cases, a different strategy to deal with the ICE may be more reasonable.

We have assumed there is only one ICE under consideration. In practice there is typically more than one ICE. With multiple types of ICE which are all chosen to be handled using the hypothetical strategy, the methods described here could be applied with the $A_1,\dots,A_K$ now denoting occurrence of at least one of the ICEs. However, in order to avoid model misspecification it may be preferable to define $A_k$ as a vector indicating occurrence or not of each of the ICE types at time $k$. If some ICE are to be dealt with using the hypothetical strategy and some using treatment policy, if the treatment policy ICE precedes the hypothetical ICE in time, it may be possible to consider the treatment policy ICE as an additional time-varying covariate (i.e. as part of $L_1,\dots,L_K$). We will address in more detail estimation in the case of two or more ICE types in a subsequent paper.



The validity of the estimates given by the methods described depends on the models being correctly specified, as shown by the simulations. To make the estimates more robust to model misspecification, so-called doubly-robust estimators were developed \citep{Bang2005DoublyRobust}. The idea is that the estimates are derived using two separate models and only one of the models needs to be correctly specified to obtain consistent estimates. The first model typically concerns the treatment assignment (propensity score model) while the second model is a model for the outcome. Section 21.3 of \cite{Hernan2020} describes calculation of a doubly robust estimator in the time-varying treatment setting, which could be applied for estimation of hypothetical estimands. To provide further robustness to model misspecification, machine learning approaches could also be explored \citep{Van2006TMLE}. 

For the methods covered in the paper, we did not discuss variance estimation. First, the purpose of the paper was to provide feasible ways to implement existing estimators to target hypothetical estimands, so the main focus was on the treatment effect estimate. More importantly, as these are well established estimators, there is already existing literature proposing different ways of estimating the corresponding variance, including bootstrapping and sandwich estimators. In fact, there are different packages available to implement them in standard software \citep{McGrath2020GformulaR,Van2011ipwpackage}.

Besides the causal methods we covered, G-estimation is another method from causal inference that is sometimes used \citep{Hernan2020}. The difference between G-estimation and the other G-methods is that G-estimation estimates conditional treatment effects within levels of $L$. In settings where the causal effect of interest is defined for a particular $L$ stratum, G-estimation may be of relevance.

The ICH E9 estimand addendum highlights the difference between missing data arising due to an ICE and data missing due to other reasons e.g. lost to follow-up. As we have discussed, certain estimators do not use data after the occurrence of an ICE, and so missingness in variables at follow-up visits occurring after ICEs present no difficulties for these estimators. Patients lost to follow-up prior to an ICE occurring would however require imputation (implicitly or explicitly) of post drop out variables, including variables indicating future occurrence of ICE. In some settings, e.g. where the ICE corresponds to receipt of rescue treatment, dropout might however preclude future occurrence of ICE. While plausible assumptions regarding missingness mechanisms and the best method to handle missingness would need to be judged on a case by case basis, MI seems attractive, given its widespread implementation in software and its ability to handle missingness in variables of mixed type. Here MI could first be used to generate completed datasets, following which any of the methods described previously (e.g. G-formula or IPW) could be applied.


We hope that by drawing parallels between causal inference and missing data methods and describing how the different estimators work, the assumptions required for valid estimates and showing feasible ways to implement them, researchers involved in the design and analysis of clinical trials will be able to successfully apply these methods to their trials.


\bigskip
\begin{center}
{\large\bf SUPPLEMENTARY MATERIAL}
\end{center}

\section*{Equivalence of G-formula and likelihood based missing data approaches}

Following Figure \ref{fig:SWIG_twotimes}, consider the data on $(L_0,L_1,L_2,Y)$ in those with $A_{0}=a_{0}$ resulting after deletion of the $L_2$ and $Y$ values for individuals with $A_1=1$ and deletion of the $Y$ values for those with $A_1=0$ and $A_2=1$. Suppose we assume the `full data' on $(L_1,L_2^{a_1=0},Y^{a_1=0,a_2=0})$ are tri-variate normal, with their means depending on $L_0$ linearly with distinct coefficients and an unstructured covariance matrix. Then viewing $(L_1,L_2^{a_1=0},Y^{a_1=0,a_2=0})$ as repeated measures, as per the case in Section \ref{sec:singletime}, this implies that the conditional means of each `outcome' given the earlier values (and $L_0$) are
\begin{align*}
    E(Y^{a_1=0,a_2=0} |A_0=a_0,L_0, L_1,L_2^{a_1=0}) &= \beta_{30}^{a_0} + \beta_{31}^{a_0} L_{0} + \beta_{32}^{a_0} L_{1} + \beta_{33}^{a_0} L_2^{a_1=0} \nonumber \\
    E(L_2^{a_1=0} |A_0=a_0, L_0, L_1) &= \beta_{20}^{a_0} + \beta_{21}^{a_0} L_{0} + \beta_{22}^{a_0} L_{1} \\
    E(L_1|A_{0}=a_{0},L_{0}) &=\beta_{10}^{a_0} + \beta_{11}^{a_0} L_{0} 
\end{align*}
Moreover, under MAR the observed data MLEs of the parameters in these models are obtained by fitting the model for $Y^{a_1=0,a_2=0}$ in those with $A_1=0$ and $A_2=0$, the model for $L_2^{a_1=0}$ in those with $A_1=0$ and the model for $L_1$ using all patients with $A_0=a_0$. Then we have that
\begin{align*}
    E(Y^{a_1=0,a_2=0} |A_0=a_{0},L_0, L_1) &= \beta_{30}^{a_0} + \beta_{31}^{a_0} L_{0} + \beta_{32}^{a_0} L_{1} + \beta_{33}^{a_0} (\beta_{20}^{a_0} + \beta_{21}^{a_0} L_{0} + \beta_{22}^{a_0} L_{1}) \\
    E(Y^{a_1=0,a_2=0} |A_0=a_{0},L_0) &= \beta_{30}^{a_0} + \beta_{31}^{a_0} L_{0} + \beta_{32}^{a_0} (\beta_{10}^{a_0} + \beta_{11}^{a_0} L_{0}) + \beta_{33}^{a_0} (\beta_{20}^{a_0} + \beta_{21}^{a_0} L_{0} + \beta_{22}^{a_0} (\beta_{10}^{a_0} + \beta_{11}^{a_0} L_{0})) \\
    E(Y^{a_1=0,a_2=0} |A_0=a_{0}) &= \beta_{30}^{a_0} + \beta_{31}^{a_0} E(L_0|A_0=a_0) + \beta_{32}^{a_0} (\beta_{10}^{a_0} + \beta_{11}^{a_0} E(L_0|A_0=a_0)) \\
    & + \beta_{33}^{a_0} (\beta_{20}^{a_0} + \beta_{21}^{a_0} E(L_0|A_0=a_0) + \beta_{22}^{a_0} (\beta_{10}^{a_0} + \beta_{11}^{a_0} E(L_0|A_0=a_0)))
\end{align*}
As in Section \ref{sec:singletime}, the non-parametric MLE of $E(L_0|A_0=a_0)$ is $\hat{E}(L_{0}|A_{0}=a_{0})=\frac{\sum^{n}_{i=1} I(A_{0,i}=a_0) L_{0,i}}{\sum^{n}_{i=1} I(A_{0,i}=a_0)}$. Then the MLE of $E(Y^{a_1=0,a_2=0}|A_0=a_0)$ is
\begin{align*}
    \hat{E}(Y^{a_1=0,a_2=0} |A_0=a_{0}) &= \hat{\beta}_{30}^{a_0} + \hat{\beta}_{31}^{a_0} \hat{E}(L_0|A_0=a_0) + \hat{\beta}_{32}^{a_0} (\hat{\beta}_{10}^{a_0} + \hat{\beta}_{11}^{a_0} \hat{E}(L_0|A_0=a_0)) \\
    & + \hat{\beta}_{33}^{a_0} (\hat{\beta}_{20}^{a_0} + \hat{\beta}_{21}^{a_0} \hat{E}(L_0|A_0=a_0) + \hat{\beta}_{22}^{a_0} (\hat{\beta}_{10}^{a_0} + \hat{\beta}_{11}^{a_0} \hat{E}(L_0|A_0=a_0))) \\
     &= \hat{\beta}_{30}^{a_0} + \hat{\beta}_{31}^{a_0} \hat{E}(L_0|A_0=a_0) + \hat{\beta}_{32}^{a_0} \hat{E}(L_1|A_0=a_0) \\
    & + \hat{\beta}_{33}^{a_0} (\hat{\beta}_{20}^{a_0} + \hat{\beta}_{21}^{a_0} \hat{E}(L_0|A_0=a_0) + \hat{\beta}_{22}^{a_0} \hat{E}(L_1|A_0=a_0)) \\
    &= \frac{\sum^{n}_{i=1} I(A_{0,i}=a_0) \{\hat{\beta}_{30}^{a_0} + \hat{\beta}_{33}^{a_0} \hat{\beta}_{20}^{a_0} + (\hat{\beta}_{31}^{a_0} + \hat{\beta}_{33}^{a_0} \hat{\beta}_{21}^{a_0}) L_{0,i} + (\hat{\beta}_{32}^{a_0} + \hat{\beta}_{33}^{a_0} \hat{\beta}_{22}^{a_0}) L_{1,i} \} }{\sum^{n}_{i=1} I(A_{0,i}=a_0)}
\end{align*}
which is the G-formula estimator of equation \eqref{eq:gform_multtime_linear}.

The key difference between the g-formula and standard regression in settings with time-varying treatment such as the one outlined in this paper is that g-formula in general fits models sequentially, to each $L_k$ ($k>0$) and $Y$  given the past, and effectively predicts (or more generally simulates) each time-dependent confounder, $\tilde{L}_k^{a_0,\ldots,a_{k-1}}$, under each treatment regime to be compared. When estimating the mean potential outcome at the final time-point under a given regime, the predictions for $Y$ that are averaged are based on the fitted model for $Y$ given the past, but with the treatments set to their values under the regime, and the confounders set to their predicted (or more generally simulated) values, again under the relevant regime. For example, if we write $m(a_0,a_1,a_2,l_0,l_1,l_2)$ for the estimated prediction function corresponding to $E(Y|a_{0},a_1,a_2,l_0,l_1,l_2,)$ then $E(Y^{a_0=a_1=a_2=0})$ would be estimated as
$$\frac{1}{n}\sum_{i=1}^n m(0,0,0,L_{0,i},\tilde{L}_{1,i}^{a_0=0},\tilde{L}_{2,i}^{a_0=a_1=0}) $$
but 
$E(Y^{a_0=1,a_1=a_2=0})$ would be estimated as
$$\frac{1}{n}\sum_{i=1}^n m(1,0,0,L_{0,i},\tilde{L}_{1,i}^{a_0=1},\tilde{L}_{2,i}^{a_0=1,a_1=0}).$$
Crucial to the success of the above strategy for correctly allowing for time-dependent confounding affected by treatment is that, in general, $\tilde{L}_{1,i}^{a_1=0}\neq \tilde{L}_{1,i}^{a_1=1}$, and so on.

How is it possible, therefore, for standard missing data methods to be equivalent to g-formula in the setting considered in this paper, even though the missing data approach appears to perform the estimation (which can usefully for the purposes of this paragraph be thought of as an imputation-based procedure) under \emph{only one} regime? There are three components to the answer to this question. First of all, we are only interested in two regimes, namely $(1,0,0)$ and $(0,0,0)$: active vs.\ control with all ICEs prevented. Second, the initial treatment is randomised (there is no confounding by $L_0$), and thus instead of needing to simulate $L_1$, $L_2$ and $Y$ based on $L_0$ for \emph{all} participants under the two regimes $(1,0,0)$ and $(0,0,0)$ before standardising to the overall distribution of $L_0$, we can simulate $L_1$, $L_2$ and $Y$ once for each participant under the regime $(A_0,0,0)$ and take the average within each treatment arm separately as estimates of $E(Y^{1,0,0})$ and $E(Y^{0,0,0})$. In an observational study, such a strategy would be biased due to the unaccounted confounding by $L_0$, but in an (initially) randomised trial, imputing separately for each arm of the trial, and thus only under one regime (`assigned treatment followed by no ICE') for each participant is valid. The third reason why the missing data approaches work is that they delete data once an ICE has occurred, so that all the remaining data are observed under the regime of interest, and hence the implied imputations (under an MAR assumption) for all the deleted data are also made under the regime of interest, exactly as would happen in the g-formula.

The previous paragraph highlights an important consideration when using missing data methods to estimate the hypothetical estimand, namely that it is not only previous measurements of the outcome that need to be deleted in those for whom an ICE has occurred, but any other time-varying common cause of $A_k$ and $Y$ that could plausibly be affected by treatment. In other words, any $L_k$ must be treated in the same way as $Y$, irrespective of whether or not it is an intermediate measurement of the final outcome of interest.






\bibliographystyle{agsm}

\bibliography{Hypothetical_estimand}

@misc{ICHE9Addendum,
author = {{International Council for Harmonisation of Technical Requirements for Pharmaceuticals for Human Use}},
title = {{Addendum on estimands and sensitivity analysis in clinical trials to the guideline on statistical principles for clinical trials E9(R1)}},
year = {2019}
}

@article{Daniel2013,
  title={Methods for dealing with time-dependent confounding},
  author={Daniel, Rhian M and Cousens, SN and De Stavola, BL and Kenward, Michael G and Sterne, JAC},
  journal={Statistics in Medicine},
  volume={32},
  number={9},
  pages={1584--1618},
  year={2013},
  publisher={Wiley Online Library}
}

@article{seaman2013review,
  title={Review of inverse probability weighting for dealing with missing data},
  author={Seaman, Shaun R and White, Ian R},
  journal={Statistical methods in medical research},
  volume={22},
  number={3},
  pages={278--295},
  year={2013},
  publisher={Sage Publications Sage UK: London, England}
}

@article{national2010prevention,
  title={The prevention and treatment of missing data in clinical trials},
  author={{National Research Council}},
  year={2010},
  publisher={National Academies Press}
}

@article{lipkovich2020causal,
  title={Causal inference and estimands in clinical trials},
  author={Lipkovich, Ilya and Ratitch, Bohdana and Mallinckrodt, Craig H},
  journal={Statistics in Biopharmaceutical Research},
  volume={12},
  number={1},
  pages={54--67},
  year={2020},
  publisher={Taylor \& Francis}
}

@article{carpenter2013analysis,
  title={Analysis of longitudinal trials with protocol deviation: a framework for relevant, accessible assumptions, and inference via multiple imputation},
  author={Carpenter, James R and Roger, James H and Kenward, Michael G},
  journal={Journal of Biopharmaceutical Statistics},
  volume={23},
  number={6},
  pages={1352--1371},
  year={2013},
  publisher={Taylor \& Francis}
}

@article{mallinckrodt2012structured,
  title={A structured approach to choosing estimands and estimators in longitudinal clinical trials},
  author={Mallinckrodt, CH and Lin, Q and Lipkovich, I and Molenberghs, Geert},
  journal={Pharmaceutical Statistics},
  volume={11},
  number={6},
  pages={456--461},
  year={2012},
  publisher={Wiley Online Library}
}

@article{mallinckrodt2020aligning,
  title={Aligning estimators with estimands in clinical trials: putting the ICH E9 (R1) guidelines into practice},
  author={Mallinckrodt, CH and Bell, J and Liu, G and Ratitch, B and O’Kelly, M and Lipkovich, I and Singh, P and Xu, L and Molenberghs, G},
  journal={Therapeutic Innovation \& Regulatory Science},
  volume={54},
  number={2},
  pages={353--364},
  year={2020},
  publisher={Springer}
}

@book{daniels2008missing,
  title={Missing data in longitudinal studies: Strategies for Bayesian modeling and sensitivity analysis},
  author={Daniels, Michael J and Hogan, Joseph W},
  year={2008},
  publisher={CRC press}
}

@article{holzhauer2015choice,
  title={Choice of estimand and analysis methods in diabetes trials with rescue medication},
  author={Holzhauer, Bj{\"o}rn and Akacha, Mouna and Bermann, Georgina},
  journal={Pharmaceutical statistics},
  volume={14},
  number={6},
  pages={433--447},
  year={2015},
  publisher={Wiley Online Library}
}

@book{Hernan2020,
  author    = "Hernan, Miguel A and Robins, James M",
  title     = "Causal Inference: What If",
  publisher = "Boca Raton: Chapman \& Hall/CRC",
  year      = "2020"
}

@article{Richardson2013SWIGs,
  title={Single world intervention graphs (SWIGs): A unification of the counterfactual and graphical approaches to causality},
  author={Richardson, Thomas S and Robins, James M},
  journal={Center for the Statistics and the Social Sciences, University of Washington Series. Working Paper},
  volume={128},
  number={30},
  pages={2013},
  year={2013}
}

@book{tsiatis2020dynamic,
  title={Dynamic Treatment Regimes: Statistical Methods for Precision Medicine},
  author={Tsiatis, Anastasios A and Davidian, Marie and Holloway, Shannon T and Laber, Eric B},
  year={2020},
  publisher={CRC press}
}

@article{Ster2020SystermaticReview,
  title={Current approaches to handling rescue medication in asthma and eczema randomised controlled trials are inadequate: a systematic review},
  author={Ster, Anca Maria Chis and Cornelius, Victoria and Cro, Suzie},
  journal={Journal of Clinical Epidemiology},
  volume={125},
  pages={148--157},
  year={2020},
  publisher={Elsevier}
}

@article{Rubin1974,
author = {Rubin, Donald B.},
doi = {10.1037/h0037350},
issn = {0022-0663},
journal = {Journal of Educational Psychology},
number = {5},
pages = {688--701},
title = {{Estimating causal effects of treatments in randomized and nonrandomized studies.}},
url = {http://content.apa.org/journals/edu/66/5/688},
volume = {66},
year = {1974}
}

@article{Bowden2020IV,
  title={Connecting Instrumental Variable methods for causal inference to the Estimand Framework},
  author={Bowden, Jack and Bornkamp, Bjoern and Glimm, Ekkehard and Bretz, Frank},
  journal={arXiv preprint arXiv:2012.03786},
  year={2020}
}

@article{Michiels2021Balanced,
  title={A novel estimand to adjust for rescue treatment in randomized clinical trials},
  author={Michiels, Hege and Sotto, Cristina and Vandebosch, An and Vansteelandt, Stijn},
  journal={Statistics in Medicine},
  year={2021},
  publisher={Wiley Online Library}
}

@article{Qu2021PrincipalStrat,
  title={Implementation of tripartite estimands using adherence causal estimators under the causal inference framework},
  author={Qu, Yongming and Luo, Junxiang and Ruberg, Stephen J},
  journal={Pharmaceutical Statistics},
  volume={20},
  number={1},
  pages={55--67},
  year={2021},
  publisher={Wiley Online Library}
}

@book{Mallinckrodt2019EstimandBook,
  title={Estimands, Estimators and Sensitivity Analysis in Clinical Trials},
  author={Mallinckrodt, Craig and Molenberghs, Geert and Lipkovich, Ilya and Ratitch, Bohdana},
  year={2019},
  publisher={CRC Press}
}

@article{Petersen2012Positivity,
  title={Diagnosing and responding to violations in the positivity assumption},
  author={Petersen, Maya L and Porter, Kristin E and Gruber, Susan and Wang, Yue and Van Der Laan, Mark J},
  journal={Statistical Methods in Medical Research},
  volume={21},
  number={1},
  pages={31--54},
  year={2012},
  publisher={Sage Publications Sage UK: London, England}
}

@article{Vanderweele2009Consistency,
  title={Concerning the consistency assumption in causal inference},
  author={VanderWeele, Tyler J},
  journal={Epidemiology},
  volume={20},
  number={6},
  pages={880--883},
  year={2009},
  publisher={LWW}
}

@article{Westreich2015Imputation,
  title={Imputation approaches for potential outcomes in causal inference},
  author={Westreich, Daniel and Edwards, Jessie K and Cole, Stephen R and Platt, Robert W and Mumford, Sunni L and Schisterman, Enrique F},
  journal={International journal of Epidemiology},
  volume={44},
  number={5},
  pages={1731--1737},
  year={2015},
  publisher={Oxford University Press}
}

@article{McGrath2020GformulaR,
  title={gfoRmula: An R package for estimating the effects of sustained treatment strategies via the parametric g-formula},
  author={McGrath, Sean and Lin, Victoria and Zhang, Zilu and Petito, Lucia C and Logan, Roger W and Hern{\'a}n, Miguel A and Young, Jessica G},
  journal={Patterns},
  volume={1},
  number={3},
  pages={100008},
  year={2020},
  publisher={Elsevier}
}

@article{erler2019jointai,
  title={JointAI: joint analysis and imputation of incomplete data in R},
  author={Erler, Nicole S and Rizopoulos, Dimitris and Lesaffre, Emmanuel MEH},
  journal={arXiv preprint arXiv:1907.10867},
  year={2019}
}

@article{Daniel2011GformulaStata,
  title={gformula: Estimating causal effects in the presence of time-varying confounding or mediation using the g-computation formula},
  author={Daniel, Rhian M and De Stavola, Bianca L and Cousens, Simon N},
  journal={The Stata Journal},
  volume={11},
  number={4},
  pages={479--517},
  year={2011},
  publisher={SAGE Publications Sage CA: Los Angeles, CA}
}

@article{Bang2005DoublyRobust,
  title={Doubly robust estimation in missing data and causal inference models},
  author={Bang, Heejung and Robins, James M},
  journal={Biometrics},
  volume={61},
  number={4},
  pages={962--973},
  year={2005},
  publisher={Wiley Online Library}
}

@article{Van2011ipwpackage,
  title={{ipw: an R package for inverse probability weighting}},
  author={van der Wal, Willem M and Geskus, Ronald B and others},
  journal={J Stat Softw},
  volume={43},
  number={13},
  pages={1--23},
  year={2011}
}

@article{Van2006TMLE,
  title={Targeted maximum likelihood learning},
  author={Van Der Laan, Mark J and Rubin, Daniel},
  journal={The International Journal of Biostatistics},
  volume={2},
  number={1},
  year={2006},
  publisher={De Gruyter}
}

@article{Muller2018Diabetes,
  title={Efficacy and safety of dapagliflozin or dapagliflozin plus saxagliptin versus glimepiride as add-on to metformin in patients with type 2 diabetes},
  author={M{\"u}ller-Wieland, Dirk and Kellerer, Monika and Cypryk, Katarzyna and Skripova, Dasa and Rohwedder, Katja and Johnsson, Eva and Garcia-Sanchez, Ricardo and Kurlyandskaya, Raisa and Sj{\"o}str{\"o}m, C David and Jacob, Stephan and others},
  journal={Diabetes, Obesity and Metabolism},
  volume={20},
  number={11},
  pages={2598--2607},
  year={2018},
  publisher={Wiley Online Library}
}

@inbook{Robins2009Chapter,
  author={Robins, James M and Hern{\'a}n, Miguel A},
  title={Longitudinal data analysis},
  chapter   = {Estimation of the causal effects of time-varying exposures},
  publisher = {CRC Press Boca Raton, FL},
  year={2009},
}
\end{document}